\documentclass[10pt,journal, final]{IEEEtran}
\usepackage{ifpdf}
\usepackage{cite}
\usepackage{algpseudocode}
\usepackage{algorithm}
\ifCLASSINFOpdf
\usepackage[pdftex]{graphicx}

\else
 \usepackage[dvips]{graphicx}
\fi
\usepackage[cmex10]{amsmath}
\usepackage{epstopdf}
\usepackage{array}
\usepackage{flushend}
\usepackage{balance}
\usepackage{eqparbox}

\usepackage[tight,footnotesize]{subfigure}
\usepackage{color}
\usepackage{float}

\usepackage{dblfloatfix}
\usepackage{url}
\usepackage{cite}
\usepackage{graphicx}
\usepackage{subfigure}
\usepackage{amsmath}
\usepackage{epstopdf}

\usepackage{amssymb}
\usepackage{color}
\usepackage[nolist]{acronym}

\begin{document}
\allowdisplaybreaks[3]
\title{Sum Rate and Worst-Case SINR Optimization in Multi-HAPS–Ground Integrated Networks}
\setlength{\columnsep}{0.21 in}

\author{Shasha~Liu,~\IEEEmembership{Student Member,~IEEE,}
        Hayssam~Dahrouj,~\IEEEmembership{Senior Member,~IEEE,}
        Abla~Kammoun,~\IEEEmembership{Member,~IEEE,}
        and
        Mohamed-Slim~Alouini,~\IEEEmembership{Fellow,~IEEE}

\thanks {
Shasha Liu, Abla~Kammoun, and Mohamed-Slim Alouini are with the Division of Computer, Electrical and Mathematical Sciences and Engineering, King Abdullah University of Science and Technology, Thuwal 23955-6900, Saudi Arabia (e-mail: shasha.liu@kaust.edu.sa, abla.kammoun@kaust.edu.sa, slim.alouini@kaust.edu.sa).

Hayssam Dahrouj is with the department of Electrical Engineering, University of Sharjah, Sharjah, UAE (e-mail: hayssam.dahrouj@gmail.com).}
}

\maketitle
\begin{abstract}
Balancing throughput and fairness promises to be a key enabler for achieving large-scale digital inclusion in future vertical heterogeneous networks (VHetNets). In an attempt to address the global digital divide problem, this paper explores a multi-high-altitude platform system (HAPS)-ground integrated network, in which multiple HAPSs collaborate with ground base stations (BSs) to enhance the users' quality of service on the ground to achieve the highly sought-after digital equity. To this end, this paper considers maximizing both the network-wide weighted sum rate function and the worst-case signal-to-interference-plus-noise ratio (SINR) function subject to the same system level constraints. More specifically, the paper tackles the two different optimization problems so as to balance  throughput and fairness, by accounting for the individual HAPS payload connectivity constraints, HAPS and BS distinct power limitations, and per-user rate requirements.
This paper solves the considered problems using techniques from optimization theory by adopting a generalized assignment problem (GAP)-based methodology to determine the user association variables, jointly with successive convex approximation (SCA)-based iterative algorithms for optimizing the corresponding beamforming vectors. One of the main advantages of the proposed algorithms is their amenability for distributed implementation across the multiple HAPSs and BSs. The simulation results particularly validate the performance of the presented algorithms, demonstrating the capability of multi-HAPS networks to boost-up the overall network digital inclusion toward democratizing future digital services.
\end{abstract}
\IEEEpeerreviewmaketitle
\section{Introduction}
\subsection{Overview}
With the advent of the fifth generation of mobile communication networks (5G) continues to deliver economic and societal benefits, attention in the research and development sector is shifting towards the sixth generation of mobile communication
networks (6G).
Such networks are anticipated to offer ubiquitous coverage, ultra-wide-area broadband access capabilities anytime and anywhere to support future Internet of Things (IoT), remote area coverage, emergency communications, ecological remote sensing, thereby providing practical solutions to the digital divide problem \cite{cui2022space,deng2025native}.
While densified terrestrial networks provide notable connectivity advantages, they prove to be inadequate to meet the constantly growing and increasingly unpredictable user demands in both spatial and temporal dimensions \cite{alam2021high}. To this end, non-terrestrial networks (NTN) emerge nowadays are strong candidates to spearhead the digital sustainability goals of 6G systems \cite{reifert2022distributed,elamassie2024fso}. More specifically, NTN connectivity from the sky comprises spaceborne (i.e., geostationary earth orbit (GEO), medium earth orbit (MEO), and low earth orbitn (LEO)
satellites) and airborne (i.e., unmanned aircraft system (UAS) \cite{deng2025two} and high altitude platform system (HAPS)) vehicles, all of which may function as relay nodes or as flying base stations. Such empowered connectivity landscape gives NTN the prospects to achieve digital inclusion (i.e., connecting the unconnected and superconnecting the connected), owing to their ability to provide reliable service availability across wide areas and effectively address the demand for seamless connectivity anytime and anywhere \cite{rinaldi2020cooperative,alzenad2019coverage}.
\par
Given their favorable channel conditions, almost stationary positions, powerful capabilities, reduced round-trip delay, and ease of deployment and maintenance, HAPS promises to be the key component of NTN \cite{kurt2021vision, qiu2019air, cao2018airborne}. Specifically, positioned at 18-21 km altitude, HAPS operates at lower altitudes than satellites, resulting in a favorable link budget and improved signal-to-interference-plus-noise ratio (SINR) performance \cite{grace2011broadband}. This altitude also allows HAPS to remain almost stationary, significantly reducing doppler shift effects. HAPS systems are further able to accomodate multiple antennas, which is well-suited for multiple-input multiple-output (MIMO) system.
\par
In spite of such advantages, a single stand-alone HAPS proves to be insufficient to meet the demands of the ongoing advancements in wireless communication designs and emerging use cases.
In particular, HAPS mega-constellation is envisioned to be a potential approach to achieve data analysis platforms, high capacity, ubiquitous connectivity, and computation offloading. Specifically, HAPS mega-constellations can provide offloading heavy computations for cargo delivery, monitoring the movement
of a swarm of unmanned aerial vehicles (UAV), and edge intelligence\cite{kurt2021vision,deng2025distributed}. HAPS is designed to meet a variety of communication needs, including enhanced mobile broadband (eMBB) connectivity and ultra-reliable low-latency communication (URLLC) \cite{kurt2021vision}. In addition, to reduce the dependence on ground-based and satellite networks, HAPS is expected to provide wireless communication services and fast Internet access.
Owing to its lower latency compared to emerging satellite networks, HAPS is capable of directly serving ground users so as to augment the terrestrial networks operation \cite{hoshino2019study}.
 While ground base stations (GBSs) are primarily responsible for meeting average user demands, their coverage capabilities are fundamentally constrained by geographical and environmental limitations. As a result, certain remote and difficult-to-access regions, such as oceans, glaciers, and mountainous terrains, often remain underserved or entirely disconnected by conventional terrestrial networks. In such cases, HAPS offers a viable and effective solution by providing broadband connectivity with significantly lower latency compared to satellite networks, owing to their closer proximity to the Earth’s surface.
Beyond addressing the connectivity challenges in such regions, HAPS is designed to manage sudden or temporary surges in user traffic. These surges may occur during large public events, natural disasters, or emergency situations where rapid deployment and scalable network capacity are required. Furthermore, HAPS can enhance the performance of terrestrial networks in densely populated urban areas by offloading traffic and mitigating congestion  \cite{hsieh2020uav, xing2021high}.
 \par
To best capture such an interplay between HAPS promising data throughputs and fairness, this paper considers a multi-HAPS ground integrated network, comprising multiple HAPSs and several ground BSs. Each HAPS is composed of three key subsystems, including, a communication payload system, an energy management subsystem, and a flight control system \cite{kurt2021vision,liu2023joint}.
This design framework imposes further restrictions on its connectivity capabilities.
Therefore, the paper addresses such unique aspects of HAPS by incorporating the connectivity constraints of the multiple HAPSs' payloads into the adopted optimization frameworks. The paper then addresses two distinct optimization problems: the minimum SINR maximization problem and the weighted sum rate maximization problem. Specifically, this paper concentrates on optimizing the strategy for associating users with either BSs or HAPSs, along with determining the corresponding beamforming vectors, while adhering to the HAPSs payload connectivity limitations, the minimum rate requirement for each user and the power constraints of both HAPSs and BSs, so as to attain a democratized, digitally inclusive connectivity landscape.
\subsection{Related Work}
The management of antenna beams and on-board radio power in integrated HAPS-ground networks is essential for interference control, directly impacting the SINR levels, and consequently, key system design metrics related to throughput and spectral efficiency.
This paper primarily investigates interference management strategies, including user scheduling and beamforming for multi-HAPS-ground integrated networks.
The problem addressed in this paper is, therefore, related to the recent advances in HAPS optimization, as well as the general framework of resource allocation in both terrestrial and non-terrestrial networks.  In fact, effective resource and interference management techniques are vital to the performance of NTN systems, while service providers typically assess this performance through metrics like minimum SINR and throughput \cite{kurt2021vision}.
\par
The seminal work \cite{yang1998optimal} explores the max-min weighted SINR optimization problem with a single power constraint in a single-input single-output (SISO) network, using an extended coupling matrix for formulation and solution.
The max-min SINR problem is subsequently extended to multiple-input single-output (MISO) and MIMO scenarios in \cite{wiesel2005linear,tan2011maximizing,cai2011max,cai2012max}. Specifically, the optimization of the worst-case SINR under a predefined power constraint is addressed in \cite{wiesel2005linear}, where the problem is reformulated as a standard generalized eigenvalue problem (GEVP), using established conic optimization techniques. However, due to the relatively high computational complexity of the method proposed in \cite{wiesel2005linear}, reference \cite{tan2011maximizing} introduces an elegant algorithm for solving the max-min SINR problem that eliminates the need for eigenvector computation.
The method in \cite{tan2011maximizing} draws inspiration from the distributed power control (DPC) algorithm and utilizes nonlinear Perron-Frobenius theory. Building on this, in references \cite{cai2011max,cai2012max}, nonlinear Perron-Frobenius theory is further applied to address the max-min weighted SINR problem with multiple weighted-sum power constraints, leading to the derivation of closed-form optimal solutions.
\par
However, multi-tier networks face significant challenges, with multi-mode interference being the most critical issue.
To enhance the design and performance assessment of future wireless networks, it is essential to implement sophisticated interference mitigation approaches, including interference alignment, radio resource management, and user association \cite{hossain2014evolution}.
Reference \cite{shamsabadi2024enhancing} solves the max-min fairness problem by adopting reformulation
linearization techniques (RLT) for user scheduling problem and successive convex approximation (SCA) for  the corresponding beamforming vectors in the standalone HAPS-ground networks.
Additionally, the authors in \cite{shamsabadi2022handling} propose an iterative algorithm for optimizing subcarrier allocation and transmit power distribution to maximize the minimum SINR in vertical heterogeneous networks (VHetNets) comprising a standalone HAPS. Reference \cite{tan2011maximizing} proposes an analog beamforming algorithm aimed at maximizing the minimum SINR with low computational complexity in a HAPS communication network.
\par
Another facet of the current paper focuses on optimizing the sum rate objective, a subject that has been thoroughly examined in recent scholarly literature.
Such a problem is non-convex and NP-hard under interference networks scenarios, even in the single-antenna case \cite{luo2008dynamic}. As a result, most current research efforts focus on finding efficient, high-quality suboptimal solutions. For example, the authors in \cite{shi2011iteratively} determine the linear beamforming solution by iteratively minimizing the weighted mean-square error (MSE) to maximize the sum rate, a method commonly referred to as weighted minimum mean squared error (WMMSE). Further, reference \cite{yu2013multicell} optimizes the weighted sum rate through the joint optimization of user scheduling, beamforming, and power allocation.
Specifically, beamforming is designed using zero-forcing (ZF), while power allocation is directly optimized through a modified Newton’s method, with the user scheduling scheme held fixed.
\par
Further, several studies in the recent literature propose resource management techniques for maximizing the sum rate of non-terrestrial networks, especially for networks involving HAPS connectivity. For instance, reference \cite{lin2019robust} introduces a robust beamforming scheme for integrated satellite-HAPS network, aiming to achieve a pareto optimal balance between two competing objectives, namely, minimizing the total transmit power and maximizing the sum rate, while meeting users' quality of
service (QoS) constraints.
The study in \cite{ibrahim2015using} optimizes the power, subchannels, and time slots to maximize the sum rate in the orthogonal frequency division multiple access (OFDMA)-based HAPS network. The work in \cite{liu2023resource} adopts non-orthogonal multiple access (NOMA) and aims at maximizing sum rate as constrained by QoS demands, HAPS transmission power, and limited connectivity per NOMA group. References \cite{liu2023joint,alghamdi2024equitable} assess the role of HAPS in promoting digital inclusion and enhancing the overall sum rate within the integrated HAPS-ground network. In the hybrid satellite-HAPS-ground network, the authors in \cite{liu2023joint} aim to maximize the system's sum rate by jointly optimizing user association and beamforming vectors in an iterative approach. Specifically, user association is determined by integrating integer linear programming (ILP) with generalized assignment problem (GAP) based methods, whereas beamforming is refined using the WMMSE approach.
Reference \cite{alghamdi2024equitable} optimizes user scheduling and beamforming vectors through sparse beamforming via reweighted $\ell_0$-norm approximation and fractional programming (FP)
to enhance the sum-of-log of the long-term average rate and sum rate for the cloud-enabled HAPS-ground network.
Authors in \cite{azizi2023ris,azizi2024exploring} investigate mobility-enabled HAPS and emphasize the role of Reconfigurable Intelligent Surfaces (RIS) in aerodynamic HAPS systems. Specifically, \cite{azizi2023ris} formulates a multi-objective optimization framework to jointly design the RIS phase shifts for maximizing the cascaded channel gain, minimizing the upper bound of delay spread, and eliminating Doppler spread. Furthermore, \cite{azizi2024exploring} proposes a HAPS-RIS-assisted architecture to support UAV networks by expanding coverage and reducing the number of UAVs, while satisfying user rate constraints. However, both works mainly focus on aerial nodes and do not consider the coexistence or coordination with ground stations.  
\par
The above studies are limited to scenarios involving ground BSs and/or a single HAPS, which confine their connectivity-from-the-sky prospects. Another approach in HAPS development is, therefore, related to scenarios empowered by multiple HAPSs, e.g., references \cite{zong2012deployment, marriott2020trajectory, marriott2018energy, alsharoa2020improvement, lee2022spectral}. In particular, the authors in \cite{zong2012deployment} investigate the deployment of multiple HAPSs while accounting for the QoS requirements of ground users. Reference \cite{zong2012deployment} proposes a self-organizing game theory model in which multiple HAPSs are represented as rational and self-organizing players. The objective in \cite{zong2012deployment} is then to achieve an optimal configuration of HAPSs that maximizes the QoS of ground users. The studies \cite{marriott2020trajectory,marriott2018energy}, on the other hand, focus on optimizing the trajectories of several HAPSs by minimizing energy consumption in \cite{marriott2020trajectory} and maximizing energy storage in \cite{marriott2018energy}, respectively.
The work in \cite{alsharoa2020improvement} optimizes satellite-HAPS-ground network throughput in terms of backhaul capacity, power allocation, user association, and multi-HAPS location. In addition, considering the limited wireless backhaul for HAPS, reference \cite{lee2022spectral} examines dual-hop mixed radio-frequency/free-space optical (RF/FSO) multi-HAPS-ground networks and optimizes the association between HAPS and BSs, power allocation for both uplink and downlink, as well as the deployment and altitude of HAPS. The goal in \cite{lee2022spectral} is to maximize the sum rate, encompassing downlink and uplink rates, employing an iterative algorithm with introduced auxiliary functions.
\subsection{Contributions}
In contrast to the previous works, an integrated multi-HAPS-ground network is investigated in this paper, where service is provided to ground users through the cooperation of HAPSs and ground BSs, each equipped with multiple antennas.
Each ground user, equipped with a single antenna, has the flexibility to associate with either a HAPS or a ground BS.
The paper subsequently addresses two distinct mixed discrete-continuous optimization problems: the maximization of the minimum SINR and the maximization of the weighted sum-rate, both subject to constraints on payload, power, and minimum rate requirements.
The general framework to solve both problems is then to determine the user association problem using a GAP-based solution. With the user association fixed, we then adopt first order Taylor expansion to convexify some of the non-convex constraints, followed by the SCA-based iterative algorithm to solve the approximate problem to determine the beamforming vectors associated with the served users for each problem.
The main contributions of this paper can then be summarized as follows:
\begin{itemize}
    \item A multi-HAPS-ground integrated network consisting of multiple HAPSs and ground BSs is proposed to collaboratively serve ground users. This paper focuses on scenarios involving multiple HAPSs to highlight their potential in providing a democratized, digitally equitable connectivity platform.
    \item By refining the user scheduling approach and optimizing the corresponding beamforming vectors, the paper focuses on managing the emerging multi-mode interference encountered by both the air segment and the terrestrial segment of the considered multi-HAPS VHetNets. To this end, the paper first formulates the minimum SINR maximization problem and then addresses this complex non-convex mixed-integer optimization problem using GAP to determine the user association strategy and SCA-based iterative approach to obtain the beamforming vectors.
    \item In addition to addressing the minimum SINR maximization problem, this paper additionally investigates the problem of maximizing the overall weighted sum rate with per-user rate requirements. To solve this problem, a feasible initial solution for optimizing the beamforming vectors is obtained by solving the minimum SINR maximization problem. Building on this feasibility step, the corresponding beamforming vectors are then designed using a SCA-based approach.
    \item The paper illustrates the implementation of the proposed algorithms in a decentralized manner across multiple HAPSs and BSs through the exchange of the proper interference-related information.
    \item The paper investigates the numerical impact of augmenting the ground network with multiple HAPSs, by highlighting the trade-off between throughput and fairness resulting from the considered multi-HAPS network. The simulations specifically showcase the numerical capabilities of the integrated multi-HAPS-ground network in achieving the highly sought-after key value indicators (KVIs) of 6G systems, especially those related to the digital inclusion prospect.
\end{itemize}
\par
The remainder of the paper is structured as follows. Section II presents the system model and defines the problem constraints. Section III elaborates on the problem formulation and introduces the proposed algorithms. Section IV discusses the simulation results, showcasing the numerical performance of the proposed solution. Finally, Section V concludes the paper.

\section{System Model}
\begin{figure}[h]
\centering
\includegraphics[width=3.5in]{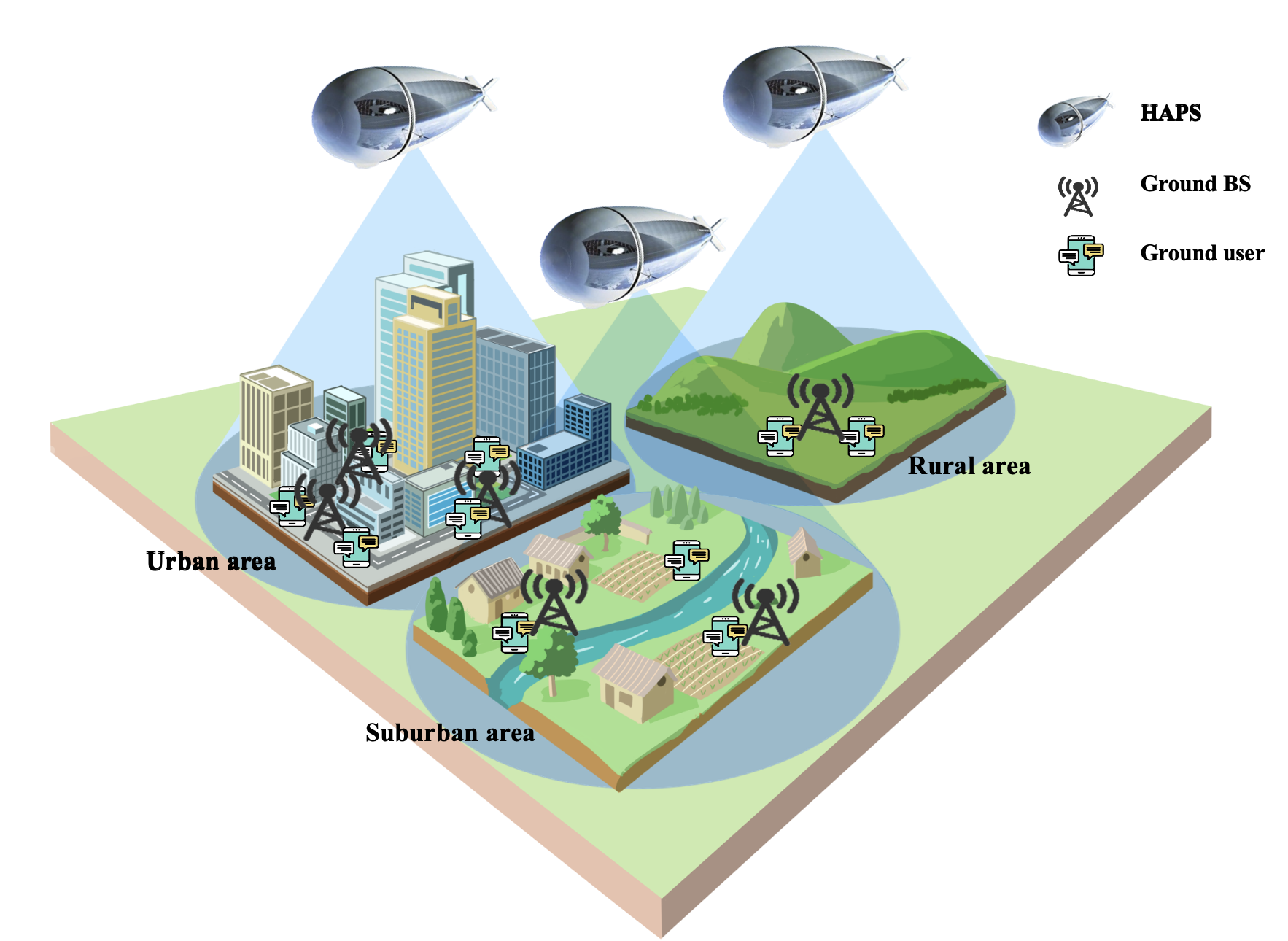}
\caption{An integrated multi-HAPS-ground network. }
\label{Fig: SM}
\end{figure}

This paper considers a VHetNet system, which consists of $M$ HAPSs, $N_B$ ground BSs, and $N_U$ users. Each HAPS is equipped with $N_{A}^{\text{HAPS}}$ antennas, and each ground BS is equipped with $N_{A}^{\text{BS}}$ antennas.
Users are indexed by $\mathcal{U}=\{1,\cdots, N_U\}$, HAPSs are indexed by $\mathcal{I}_{\text{HAPS}}=\{1,\cdots, M \}$, and ground BSs are indexed by $\mathcal{I}_{\text{BS}}=\{1, \cdots, N_B\}$. We denote the total set of transmitters (i.e., the HAPSs and BSs) by $\mathcal{I}=\mathcal{I}_{\text{HAPS}} \cup \mathcal{I}_{\text{BS}}$.
We, therefore, denote that each transmitter is equipped $N_A^i$ antennas, $\forall i \in \cal{I}$.
An example of the proposed network, comprising three HAPSs, each capable of providing service to users across urban, suburban, and rural regions, is illustrated in Fig.~\ref{Fig: SM}.
In our system, the gateway (GW) is responsible for delivering data to the HAPS, effectively serving as a backhaul link. Reference
\cite{alzenad2018fso} shows that free-space optical (FSO) links offer a high-rate, cost-efficient, easy-to-deploy, and scalable backhaul/fronthaul solution, which is essential for 5G and beyond wireless networks. In our study, therefore, we assume that the gateway employs an FSO link, referred to as an FSO gateway (FSO-GW) \cite{alzenad2018fso}. Moreover, the FSO link is assumed to operate in separate frequency bands from the downlink transmissions, thereby preventing interference with users.
Further, we assume that the central processor (CP) can acquire perfect and instantaneous channel state information (CSI). In addition, this paper adopts idealized assumptions regarding control signaling and omits the explicit modeling of communication overhead—such as signaling for CSI acquisition, user association updates, and inter-node coordination.
We assume that CP is also available to execute the algorithms. Such a CP operation can be implemented on a high-performance computing platform or an edge server with sufficient computational resources, i.e., depending on the system architecture and real-time processing requirements.
\par
\subsection{Channel Model}
For the channel model, we consider path loss, shadowing, and fading \cite{alsharoa2020improvement}. The channel {coefficient} between the $n^{th}$ antenna of the $i^{th}$ transmitter and the $j^{th}$ user is given by
\begin{equation}
\label{RF_channel}
h_{ij,n}=\left(\frac{c}{4\pi d_{ij,n}f_{c}}\right){A_{ij,n}}F_{ij,n},
\end{equation}
where $d_{ij,n}$ is the distance between the $n^{th}$ antenna of the $i^{th}$ {transmitter} and the $j^{th}$ user, $c$ represents the speed of light, and $f_{c}$  denotes the carrier frequency.
To account for the possibility of signal obstruction in terrestrial networks, the large-scale fading component \( A_{ij,n} \) is modeled as log-normal shadowing, while the small-scale fading component \( F_{ij,n} \) follows a Rayleigh distribution. This reflects the rich-scattering and non-line-of-sight (NLoS) conditions commonly found in ground-based wireless environments.
In contrast, for the HAPS-to-user links, the small-scale fading component \( F_{0j,n} \) is modeled using a Rician distribution with a Rician factor \( \kappa_{\mathrm{HAPS}} \), due to the typically strong line-of-sight (LoS) conditions enabled by the high altitude of the HAPS. Additionally, since signal obstruction is negligible in such aerial links, the large-scale fading term \( A_{0j,n} \) is set to 1.
Throughout the remainder of the paper, we refer to the general expression for the channel vector between {transmitter} $i$ and user $j$ as $\mathbf{h}_{ij}\in \mathbb{C}^{N_{A}^{i}}$, where  $\mathbf{h}_{ij}=[h_{ij,1}, h_{ij,2}, \cdot\cdot\cdot, h_{ij,n}, \cdot\cdot\cdot, h_{ij, N_{A}^{i}}]^{T}$. In the simulation section, the adopted channel parameters are explicitly specified to differentiate between the HAPS-to-user and ground BS-to-user scenarios.

\subsection{User Association Constraint}
It is assumed that each user is associated exclusively with either a HAPS or a BS, but not by both simultaneously in this paper. Therefore, we define the binary user association variable as $\alpha_{ij}$.
That is, if user $j$ is associated with transmitter $i$, then $\alpha_{ij}=1$; otherwise $\alpha_{ij}=0$.

To ensure stable operation, a HAPS platform must be equipped with several essential subsystems, including a flight control unit for mobility management and quasi-stationary positioning, an energy management system for efficient power storage and distribution, and a communication payload module to support data exchange with ground and aerial entities. These practical hardware and operational requirements necessitate the consideration of payload limitations in system design.  
Furthermore, to reflect the  HAPS payload constraints, the paper assumes that each HAPS $i$ cannot serve more than a certain number of users, which is denoted by $K_i$.
\par
In addition, HAPS payload constraint considered in the paper is assumed to be fixed based on a pre-determined step that accounts for the network traffic, available computing resources, the spatial-temporal aspects of the considered system, available bandwidth, etc. In this context, the payload limitation for HAPS $i$, i.e., $K_i$, is determined in an outer loop, and is considered constant in the context of our current paper. While we acknowledge that devising an optimized mechanism to determine $K_i$ is an interesting research direction on its own, we relegate such an interesting optimization framework for our future investigation. 
\par
Given that the data of user $j$ may not be readily accessible at the transmitter $i$ ($i\in \mathcal{I}$), we introduce the data availability variable $\beta_{ij}$. Specifically, if the data for user $j$ is accessible at the transmitter $i$, $\beta_{ij}=1 $, if not, $\beta_{ij}=0$ \footnote{The paper assumes that the data availability variable $\beta_{ij}$ is known to the optimizer.}, indicating the absence of data at that transmitter.
Both considerations above yield the following constraints:
\begin{equation}
\label{Eq: C1}
   \sum_{i \in \cal{I}}\beta_{ij}\alpha_{ij}=1, \forall j \in \mathcal{U}.
\end{equation}
\begin{equation}
\label{Eq: C2}
   \sum_{j \in \mathcal{U}}\beta_{ij}\alpha_{ij} \leq K_i, \forall i \in \mathcal{I}_{HAPS}.
\end{equation}

\subsection{Rates Expressions}
A MIMO downlink system is considered, where users share the frequency band. The signal received by user $j$, denoted as $y_j$, is formulated as:
\begin{equation}
\label{Eq: received signal}
y_{j}=\sum_{b \in \mathcal{I}}\sum_{u \in \mathcal{U}}\beta_{bu}\alpha_{bu}s_{bu}\mathbf{h}_{bj}^{H}\mathbf{w}_{bu} + z_{j},
\end{equation}
where $s_{bu}$ is the transmitted signal from transmitter $b$ to user $u$, $\mathbf{w}_{bu} \in \mathbb{C}^{N_A^i \times 1}$ is the transmit beamformer vector associated with $s_{bu}$, and
$z_{j}$ denotes the additive white circularly symmetric Gaussian complex noise, with variance $\sigma^{2}$.
Then, the $\mathrm{SINR}$ of user $j$ associated with transmitter $i$ can be written as:
\begin{equation}
\label{Eq: SINR}
\gamma_{ij}=\frac{|\mathbf{h}_{ij}^{H}\mathbf{w}_{ij}|^{2}}{\sum_{u \in \mathcal{U}, u\neq j}\sum_{b \in \mathcal{I}}\beta_{bu}\alpha_{bu}|\mathbf{h}_{bj}^{H}\mathbf{w}_{bu}|^{2} + \sigma^{2}}.
\end{equation}
Therefore, the achievable data rate of user $j$ associated with transmitter $i$ can be written as:
\begin{equation}
    \label{Eq: rate}
    R_{ij}=B\log(1+\gamma_{ij}),
\end{equation}
where $B$ is the transmission bandwidth.

\section{Problem Formulations and Proposed Solutions}
This paper addresses two key problems: maximizing the minimum SINR and maximizing the throughput, through the optimization of user scheduling and beamforming vectors for users across the network under system design constraints.
In this section, we formulate each problem separately, and propose numerically feasible algorithms to address them individually. Specifically, for both problems, we first optimize the user association variables. Then we adopt first-taylor expansion to convexify the constraint and use the SCA-based iterative method to solve each problem in the sequel.

\subsection{Maximization of Minimum $\mathrm{SINR}$}
The first optimization approach of the paper focuses on maximizing the minimum $\mathrm{SINR}$ across the network while satisfying the HAPSs individual payload constraints, user association constraints and transmit power limitations. Let  $\gamma_{\text{min}}$ represent the minimum SINR across all the network users, and let $P_{i}^{\text{max}}$ be the maximal allowable power for transmitter $i$, $\forall i\in \mathcal{I}_{\text{HAPS}} \cup \mathcal{I}_{\text{BS}}$. 
Mathematically, the fundamental max-min-$\mathrm{SINR}$ formulation of this optimization problem can be expressed as:
\begin{subequations}
\label{EHMO_sinr}
\begin{eqnarray}
\label{EHMOa_sinr}
&\displaystyle\max_{\alpha_{ij},\mathbf{w}_{ij}, }& \min_{j\in\mathcal{U}} \gamma_{j}\\
\label{EHMOb_sinr}
&s.t.& \sum_{i\in \mathcal{I}}\beta_{ij}\alpha_{ij}\gamma_{ij}=\gamma_j,  \forall j \in \mathcal{U},\\
\label{EHMOc_sinr}
&&\sum_{j \in \mathcal{U}}\beta_{ij}\alpha_{ij}\mathbf{w}_{ij}^{H}\mathbf{w}_{ij}\leq P_{i}^{\text{max}},\ \forall i \in \mathcal{I}, \\
\label{EHMOd_sinr}
&&\sum_{i\in \mathcal{I}}\beta_{ij}\alpha_{ij}=1, \forall j \in \mathcal{U},\\
\label{EHMOe_sinr}
&&\sum_{j \in \mathcal{U}}\beta_{ij}\alpha_{ij} \leq K_i, \forall i \in \mathcal{I},\\
\label{EHMOf_sinr}
&& \alpha_{ij}\in \{0, 1\},\  \forall i \in \mathcal{I}, \forall j \in \mathcal{U},
\end{eqnarray}
\end{subequations}
where the optimization is over the variables $\alpha_{ij}, \mathbf{w}_{ij}$. Constraint (\ref{EHMOb_sinr}) gives the SINR for each user, while constraint (\ref{EHMOc_sinr}) restricts the overall transmit power of each transmitter to its maximum allowable power. Additionally, constraints (\ref{EHMOd_sinr})–(\ref{EHMOf_sinr}) guarantee that each user is served by a single transmitter and restrict the number of users that each transmitter can serve. 
\par
As the objective function $\min_{j\in\mathcal{U}} \gamma_{j}$ in problem \eqref{EHMO_sinr}
is non-smooth and difficult to handle directly, we introduce the auxiliary variable $\gamma_{\text{min}}=\min_{j\in\mathcal{U}} \gamma_{j}$ and replace the original objective with a new formulation that incorporates an inequality constraint:
\begin{equation}
    \gamma_j \geq \gamma_{\text{min}},  \forall j \in \mathcal{U}
\end{equation}
This reformulation leads to an equivalent problem as follows:
\begin{subequations}
\label{EHM_sinr}
\begin{eqnarray}
\label{EHMa_sinr}
&\displaystyle\max_{\alpha_{ij},\mathbf{w}_{ij}, \gamma_{\text{min}}}& \gamma_{\text{min}}\\
\label{EHMb_sinr}
&s.t.&\gamma_j \geq \gamma_{\text{min}},  \forall j \in \mathcal{U},\\
\label{EHMc_sinr}
&& \sum_{i\in \mathcal{I}}\beta_{ij}\alpha_{ij}\gamma_{ij}=\gamma_j,  \forall j \in \mathcal{U},\\
\label{EHMd_sinr}
&&\sum_{j \in \mathcal{U}}\beta_{ij}\alpha_{ij}\mathbf{w}_{ij}^{H}\mathbf{w}_{ij}\leq P_{i}^{\text{max}},\ \forall i \in \mathcal{I}, \\
\label{EHMe_sinr}
&&\sum_{i\in \mathcal{I}}\beta_{ij}\alpha_{ij}=1, \forall j \in \mathcal{U},\\
\label{EHMf_sinr}
&&\sum_{j \in \mathcal{U}}\beta_{ij}\alpha_{ij} \leq K_i, \forall i \in \mathcal{I},\\
\label{EHMg_sinr}
&& \alpha_{ij}\in \{0, 1\},\  \forall i \in \mathcal{I}, \forall j \in \mathcal{U},
\end{eqnarray}
\end{subequations}
 where the optimization now is over the variables $\alpha_{ij}, \mathbf{w}_{ij}, \text{and\quad} \gamma_{\text{min}}$. Constraints (\ref{EHMb_sinr}) and (\ref{EHMc_sinr}) ensure that the minimum SINR is maintained throughout the network.
\par
In addition, given the numerically complex structure of the problem (\ref{EHM_sinr}), the paper solves the problem using a combination of iterative and heuristic approaches, which prove to offer an efficient, feasible solution for the optimization problem (\ref{EHM_sinr}). The proposed approach first establishes the user association scheme with predefined beamforming vectors through a GAP-type solution. Subsequently, based on the user association scheme and feasible initial solution, the beamforming vectors are found through SCA-based iterative algorithms, as discussed next.
\par
\subsubsection{User Association Strategy}
\label{subsubsection: user assocaition}
We firstly fix the beamforming vectors using an maximum ratio combining (MRC) (i.e., $\mathbf{w}_{ij}= \frac{\mathbf{h}_{ij}}{\left\|\mathbf{h}_{ij}\right\|_2^2}$).
Define $\mathcal{U}(i,j)$ as an auxiliary function that measures the benefits of connecting user $j$ to transmitter $i$, which can be written as:
\begin{equation}
    \label{Fun: U(i,j)}
\mathcal{U}_{ij}=\frac{|\mathbf{h}_{ij}^{H}\mathbf{w}_{ij}|^{2}}{\sigma^2}.
\end{equation}
Then the user association problem in (\ref{EHM_sinr}) is reformulated as:

\begin{subequations}
\label{EHM3}
\begin{eqnarray}
\label{EHM3a}
&\displaystyle\max_{\alpha_{ij}}&
\sum_{i=1}^{\operatorname{card}(\mathcal{I})}\sum_{j=1}^{N_U} \beta_{ij}\alpha_{ij}\mathcal{U}_{ij}\\
\label{EHM3b}
&s.t.& (\ref{Eq: C1}), (\ref{Eq: C2}).\\
\label{EHM3c}
&& \alpha_{ij}\in \{0, 1\},\  \forall i \in \mathcal{I}, \forall j \in \mathcal{U},
\end{eqnarray}
\end{subequations}
The problem described above is a GAP, which can be solved using standard techniques such as the branch-and-bound (BnB) algorithm \cite{douik2020tutorial}. Based on the $\alpha_{ij}$ obtained above, we
can determine the group of users assigned to transmitter $i$, which is defined as:
\begin{equation}
\label{Eq: user set}
\mathcal{U}_i=\left\{j \in \mathcal{U} \mid \beta_{i j} \alpha_{i j}=1\right\} .
\end{equation}
\par
\subsubsection{Beamforming for minimum $\mathrm{SINR}$ maximization}
Given the user association strategy resulting from solving problem (\ref{EHM3}), the paper now focuses on finding the beamforming vectors that optimize problem (\ref{EHM_sinr}). Firstly, we convexify the constraint (\ref{EHMb_sinr}) and then adopt SCA to design the beamforming vectors for problem (\ref{EHM_sinr}). More precisely, for fixed user association scheme,
we now note that problem (\ref{EHM_sinr}) can be rewritten as follows:
\begin{subequations}
\label{EHMsinBF}
\begin{eqnarray}
\label{EHMsinrBFa}
&\displaystyle &\max _{\mathbf{w}_{ij}, \gamma_{\text{min}}} \gamma_{\text{min}} \\
\label{EHMsinrBFb}
&s.t.& \frac{|\mathbf{h}_{i j}^H \mathbf{w}_{i j}|^2}
{\sum_{b \in \mathcal{I}} \sum_{u \in \mathcal{U}_b}|\mathbf{h}_{b j}^H \mathbf{w}_{b u}|^2+\sigma^2 } \geq \gamma_{\text{min}}\\
\label{EHMsinrBFc}
&&  \sum_{j \in \mathcal{U}_i} \mathbf{w}_{i j}^H \mathbf{w}_{i j} \leq P_i^{\text{max}} .
\end{eqnarray}
\end{subequations}
We can note that constraint (\ref{EHMsinrBFb}) is non-convex, because of the fractional structure of the left-hand side. In the following, we address the non-convex constraint (\ref{EHMsinrBFb}) by converting it into a convex form using the first-order
Taylor approximation, as highlighted in the following part.
\par
Assume that user $j$ is allocated to transmitter $i$ (i.e., $j \in \mathcal{U}_i$). Then, equation (\ref{EHMsinrBFb}) can be further reformulated as:
\begin{equation}
\label{Eq: sinr_min}
\sigma^2+\sum_{b=1}^{\operatorname{card}(\mathcal{I})} \sum_{u \in \mathcal{U}_{b, u \neq j}}|\mathbf{h}_{b j}^H \mathbf{w}_{b u}|^2-\frac{|\mathbf{h}_{i j}^H \mathbf{w}_{i j}|^2}{\gamma_{\text{min}}} \leq 0, \forall j \in \mathcal{U}_i .
\end{equation}
Note that equation (\ref{Eq: sinr_min}) is non-convex due to the negative fraction of the norm of the signal power and $\gamma_{\text{min}}$. To address this issue, we reformulate equation (\ref{Eq: sinr_min}) into an equivalent convex form by substituting the lefthand side with its first-order Taylor series approximation. By applying an approach similar to the one utilized in \cite{reifert2022distributed}, the following step is used to convexify the constraint (\ref{Eq: sinr_min}), which is expressed as follows:
\begin{equation}
\zeta^{+}(\mathbf{w})-\zeta^{-}\left(\mathbf{w}, \gamma_{\text{min}}\right) \leq 0, \forall j \in \mathcal{U}_i,
\end{equation}
where $\zeta^{+}(\mathbf{w})^2=\sigma^2+\sum_{b \in \mathcal{I}} \sum_{u \in \mathcal{U}_b, u \neq j}|\mathbf{h}_{b j}^H \mathbf{w}_{b u}|^2$, and $\zeta^{-}=\frac{|\mathbf{h}_{(j}^H \mathbf{w}_{i j}|^2}{\gamma_{\text{min}}}$ are convex functions. Then, the first-order approximated form of the function $\zeta^{-}$can be written as:
\begin{equation}
\frac{|\mathbf{h}_{i j}^H \mathbf{w}_{i j}|^2}{\gamma_{\text{min}}} \approx \frac{2}{\hat{\gamma}_{\text{min}}} \Re\{\hat{\mathbf{w}}_{i j}^H \mathbf{h}_{i j} \mathbf{h}_{i j}^H \mathbf{w}_{i j}\}-\frac{\gamma_{\text{min}}}{(\hat{\gamma}_{\text{min}})^2}|\mathbf{h}_{i j}^H \hat{\mathbf{w}}_{i j}|^2,
\end{equation}
where $\hat{\mathbf{w}}_{ij}$ and $\hat{\gamma}_{\text{min}}$ are feasible constant values that satisfy constraints (\ref{EHMsinrBFb}). The final term on the left-hand side of the non-convex inequality (\ref{Eq: sinr_min}) is replaced by its linearized counterpart, and the approximate function for the left side of equation (\ref{Eq: sinr_min}) is expressed as:
\begin{equation}
\label{Eq: g1}
\begin{aligned}
g_1(\gamma_{\text{min}}, \hat{\gamma}_{\text{min}}, \hat{\mathbf{w}}_{i j}, \mathbf{w}_{i j})= & \sigma^2+\sum_{b \in \mathcal{I}} \sum_{u \in \mathcal{U}_b, u \neq j}|\mathbf{h}_{b j}^H \mathbf{w}_{b u}|^2 \\
& +\frac{\gamma_{\text{min}}}{(\hat{\gamma}_{\text{min}})^2}|\mathbf{h}_{i j}^H \hat{\mathbf{w}}_{i j}|^2 \\
& -\frac{2}{\hat{\gamma}_{\text{min}}} \Re\left\{\hat{\mathbf{w}}_{i j}^H \mathbf{h}_{i j} \mathbf{h}_{i j}^H \mathbf{w}_{i j}\right\}
\end{aligned}
\end{equation}
Consequently, the problem (\ref{EHMsinBF}) can be reformulated as follows:
\begin{subequations}
\label{EHMsinBF1}
\begin{eqnarray}
\label{EHMsinrBF1a}
&\displaystyle &\max _{\mathbf{w}_{ij}, \gamma_{\text{min}}} \gamma_{\text{min}} \\
\label{EHMsinrBF1b}
&s.t.& g_1(\gamma_{\text{min}}, \hat{\gamma}_{\text{min} }, \hat{\mathbf{w}}_{i j}, \mathbf{w}_{i j}) \leq 0\\
\label{EHMsinrBF1c}
&&  \sum_{j \in \mathcal{U}_i} \mathbf{w}_{i j}^H \mathbf{w}_{i j} \leq P_i^{\text{max}} .
\end{eqnarray}
\end{subequations}
The reformulated problem (\ref{EHMsinBF1}) is now a second order cone program (SOCP) problem \cite{lobo1998applications}, which can be efficiently solved using solvers like Mosek \cite{aps2019mosek,andersen2000mosek}.
The iterative numerical procedure for the solution of problem (\ref{EHMsinBF}) is subsequently outlined in Algorithm \ref{G:R_min}. In this algorithm, we carefully select the initial values from the feasible space, and solve the optimization problem defined in (\ref{EHMsinBF1}) iteratively. The iterative process continues until the relative difference in the objective function value converges within a predefined tolerance threshold $\epsilon$.
\begin{algorithm}
 \caption{Proposed minimum rate maximization algorithm}
 \label{G:R_min}
\begin{algorithmic}
\State $n=0$, $\hat{\gamma}_{\text{min}}$ and $\mathbf{\hat{w}}_{ij}$ are feasible;
\While {$\max_j|\gamma_{\text{min}}^{(n)}-\gamma_{\text{min}}^{(n-1)}|> \epsilon$}
\State solve the problem (\ref{EHMsinBF1})
\State $\{\hat{\gamma}_{\text{min}}, \mathbf{\hat{w}}_{ij}\}=\{\gamma_{\text{min}}^{(n)}, \mathbf{w}_{ij}^{(n)}\}$;
\State $n=n+1$;
\EndWhile
 \end{algorithmic}
 \end{algorithm}

\subsection{Maximization of Weighted Sum Rate}
The above optimization problem (\ref{EHM_sinr}) inherently aims at balancing the load among the network ground users, and thus results in suboptimal strategies for maximizing the network-wide weighted throughput. The paper, therefore, now focuses on maximizing the weighted sum rate instead, so as to illustrate the proper trade-off between the two inter-related, yet competing, data-related metrics, i.e., fairness and throughput.
\par
Let $R_{j, \text{min}}$ be the minimum data rate of each user $j$, $\forall j \in \mathcal{U}$, and let $a_{ij}$ represent the weight associated with the data rate of user $j$ when served by transmitter $i$. The optimization problem is subject to user association constraints, power limitations, and per-user minimum rate requirements. Accordingly, the weighted sum rate maximization problem can be formulated as follows:
\begin{subequations}
\label{EHM}
\begin{eqnarray}
\label{EHMa}
&\displaystyle\max_{\alpha_{ij},\mathbf{w}_{ij}}&
\sum_{i\in \mathcal{I}_{\text{BS}}}
\sum_{j=1}^{N_{U}}\beta_{ij}\alpha_{ij}a_{ij}R_{ij}
\!\!\!+\!\!\!\! \sum_{i\in \mathcal{I}_{\text{HAPS}}} \sum_{j=1}^{N_{U}}\beta_{ij}\alpha_{ij}a_{ij}R_{ij}
\\
\label{EHMb}
&s.t.& R_j \geq R_{j,\text{min}},\forall j \in \mathcal{U},\\
\label{EHMc}
&& \sum_{i\in \mathcal{I}}\beta_{ij}\alpha_{ij}R_{ij}=R_j\\
\label{EHMd}
&&\sum_{i\in \mathcal{I}}\beta_{ij}\alpha_{ij}=1, \forall j \in \mathcal{U},\\
\label{EHMe}
&&\sum_{j \in \mathcal{U}}\beta_{ij}\alpha_{ij} \leq K_i, \forall i \in \mathcal{I},\\
\label{EHMf}
&&\alpha_{ij}\in \{0, 1\},\  \forall i \in \mathcal{I}, \forall j \in \mathcal{U}, \\
\label{EHMg}
&& \sum_{j \in \mathcal{U}}\beta_{ij}\alpha_{ij}\mathbf{w}_{ij}^{H}\mathbf{w}_{ij}\leq P_{i}^{\text{max}},\ \forall i \in \mathcal{I},
\end{eqnarray}
\end{subequations}
where constraints (\ref{EHMb}) and (\ref{EHMc}) guarantee the minimum data rate of each user $j$. Constraint (\ref{EHMd}) imposes that every user is associated with one transmitter, while constraint (\ref{EHMe}) ensures that the payload of each HAPS $i$ cannot exceed a nominal capacity. Constraint (\ref{EHMg}) limits the total transmit power of transmitter $i\in \cal{I}$. 
\par
Once again, one can note that the above problem (\ref{EHM}) is a non-convex optimization problem involving both discrete and continuous variables. To address this, we first find the user-association strategy by adopting a similar GAP-based solution as in steps (\ref{EHM3}-\ref{Eq: user set}) above. Subsequently, we employ an iterative method based on SCA to design the corresponding beamforming vectors. The detailed approach is outlined in the  following part.
\subsubsection{User Association Strategy}
Similar to one of the steps of the earlier subsection \ref{subsubsection: user assocaition}, we adopt the GAP framework to maximize the utility function and derive the user association variable. For brevity and completeness, we note that we determine the group of users assigned to transmitter $i$, which is defined as:
\begin{equation}
\label{user_association_sum_rate}
\mathcal{U}_i=\left\{j \in \mathcal{U} \mid \beta_{i j} \alpha_{i j}=1\right\} .
\end{equation}
\subsubsection{Beamforming for weighted sum rate maximization}
In this subsection, we focus on optimizing the beamforming vectors for problem (\ref{EHM}). We, therefore, utilize the first-order Taylor series expansion to approximate the non-convex per-user minimum rate constraints (\ref{EHMb}) and (\ref{EHMc}), thereby transforming problem (\ref{EHM}) into a mathematically tractable form. Given the user association variables found through the set $\mathcal{U}_i$ in (\ref{user_association_sum_rate}), the problem (\ref{EHM}) can be rewritten as:
\begin{subequations}
\label{EHM1}
\begin{eqnarray}
\label{EHM1a}
&\displaystyle\max_{\mathbf{w}_{ij}}&
\sum_{j \in \mathcal{U}} \hat{a}_{j}R_j
\\
\label{EHM1b}
&s.t.& {\color{black}{R_j \geq R_{j,\text{min}},\forall j \in \mathcal{U}}},\\
\label{EHM1c}
&& \sum_{i\in \mathcal{I}}\beta_{ij}\alpha_{ij}R_{ij}=R_j\\
\label{EHM1d}
&& \sum_{j \in \mathcal{U}}\beta_{ij}\alpha_{ij}\mathbf{w}_{ij}^{H}\mathbf{w}_{ij}\leq P_{i}^{\text{max}},\ \forall i \in \mathcal{I},
\end{eqnarray}
\end{subequations}
$\hat{a}_j=\sum_{i\in \mathcal{I}}\beta_{ij}\alpha_{ij}a_{ij}$is the effective weight associated with the rate of user $j$.
\par
To best handle the non-convexity of constraint (\ref{EHM1b}), we introduce an auxiliary variable $\bar{\gamma}_{j}$ to replace the SINR term of user $j$ associated with its corresponding transmitter. In addition, Assume that user $j$ is allocated to transmitter $i$ (i.e., $j\in\mathcal{U}_{i}$).
Then the per-user rate constraint (\ref{EHM1b}) can be rewritten as follows:
\begin{equation}
\label{Eq:cons_R_min1}
 B\log(1+\bar{\gamma}_{j}) \geq R_{j,\text{min}}, \forall j,
\end{equation}
and
\begin{equation}
\label{Eq:cons_R_min2}
 \frac{|\mathbf{h}_{ij}^{H}\mathbf{w}_{ij}|^{2}}{\sum_{b=1}^{\operatorname{card}(\mathcal{I})}\sum_{u\in \mathcal{U}_{b},u \neq j}|\mathbf{h}_{bj}^{H}\mathbf{w}_{bu}|^{2}+ \sigma^{2}} \geq \bar{\gamma}_j.
\end{equation}
We can first linearize equation (\ref{Eq:cons_R_min1}) as follows:
\begin{equation}
\label{Eq: R_min_constraint}
    \bar{\gamma}_{j} \geq \gamma_{j,min}.
\end{equation}
where $\gamma_{j,\text{min}}=2^{R_{j,min}/B}-1$. Then, equation (\ref{Eq:cons_R_min2}) can be further reformulated as:
\begin{equation}
\label{Eq:cons_R_min21}
\sigma^{2} + \sum_{b=1}^{\operatorname{card}(\mathcal{I})}\sum_{u\in \mathcal{U}_{b},u \neq j}|\mathbf{h}_{bj}^{H}\mathbf{w}_{bu}|^{2}-\frac{|\mathbf{h}_{ij}^{H}\mathbf{w}_{ij}|^{2}}{\bar{\gamma}_{j}} \leq 0, \forall j\in\mathcal{U}_{i}.
\end{equation}
Note that equation (\ref{Eq:cons_R_min21}) is non-convex.
In contrast to the previous equation (\ref{Eq: sinr_min}), however, the denominator of the fraction in (\ref{Eq:cons_R_min21}) involves the auxiliary variable $\bar{\gamma}_j$ rather than $\gamma_{\text{min}}$. This emerging distinction occurs because $\bar{\gamma}_j$ is now used to replace the $\mathrm{SINR}$ of user $j$ in (\ref{Eq:cons_R_min21}).
By employing a method similar to the one used in previous subsection to convexify the constraint (\ref{Eq:cons_R_min21}), we get:
\begin{eqnarray}
    \label{Fun:g2}
    g_2(\bar{\gamma}_{j}, \hat{\gamma}_{j},\mathbf{\hat{w}}_{ij}, \mathbf{w}_{ij} )&\!\!\!=\!\!\!&\sigma^{2} + \sum_{b \in \mathcal{I}}\sum_{u\in \mathcal{U}_{b}, u\neq j}|\mathbf{h}_{bj}^{H}\mathbf{w}_{bu}|^{2} \nonumber\\
    & &
    + \frac{\bar{\gamma}_{j}}{(\hat{\gamma}_{j})^2} |\mathbf{h}_{ij}^{H}\mathbf{\hat{w}}_{ij}|^{2} \nonumber\\
    & &
    - \frac{2}{\hat{\gamma}_{j}}\Re\left\{\mathbf{\hat{w}}_{ij}^{H}\mathbf{h}_{ij} \mathbf{h}_{ij}^{H}\mathbf{w}_{ij} \right\},
\end{eqnarray}
where $\mathbf{\hat{w}}_{ij}$ and $\hat{\gamma}_{j}$ are feasible fixed values satisfying constraints (\ref{EHM1b}).
Therefore, we can conclude that equation  (\ref{EHM1b}) can be replaced by the following two new constraints as below:
\begin{subequations}
\begin{eqnarray}
      2^{R_{j,min}/B}-1-\bar{\gamma}_j &\leq & 0,  \\
      g_2(\gamma_{j}, \hat{\gamma}_{j},\mathbf{\hat{w}}_{ij}, \mathbf{w}_{ij} )&\leq & 0.
\end{eqnarray}
\end{subequations}
Notably, by applying the aforementioned relaxations, we can convexify the per-user minimum rate constraints. 
However, under the given user association strategy, if the SINR threshold for user is set too high, the optimization problem defined in equation~\eqref{EHM1} may become infeasible. Moreover, finding a feasible solution for \( \hat{\gamma}_{j} \) and \( \mathbf{\hat{w}}_{ij} \) that satisfies constraint~\eqref{EHM1b} becomes particularly challenging in scenarios characterized by high interference and stringent SINR requirements.
\par
To address these issues, we introduce a preprocessing step prior to the beamforming optimization. Specifically, we formulate a minimum SINR maximization problem for all users, as given in equation~\eqref{EHMsinrBFa}, which serves two purposes: it provides a reference for setting appropriate SINR thresholds and yields a feasible initial point for the subsequent beamforming design. This step is essential to ensure the feasibility of the overall optimization problem and to enhance the robustness of the proposed algorithm in interference-limited environments.
\par
 By applying the per-user minimum rate convexification technique, i.e., by calling the routine of Algorithm \ref{G:R_min} above, and by finding the feasible solution through such a feasibility step, the beamforming optimization problem is now formulated as follows:
\begin{subequations}
\label{EHM2}
\begin{eqnarray}
\label{EHM2a}
&\displaystyle\max_{\mathbf{w}_{ij},\bar{\gamma}_j}&
\sum_{j\in \mathcal{U}}
\hat{a}_jB\log_2(1+\bar{\gamma}_j)\\
\label{EHM2b}
&s.t.&
2^{R_{j,min}/B}-1-\bar{\gamma}_j \leq  0 \\
\label{EHM2c}
&& g_2(\bar{\gamma}_{j}, \hat{\gamma}_{j},\mathbf{\hat{w}}_{ij}, \mathbf{\hat{w}}_{ij}) \leq 0,\\
\label{EHM2d}
&&\sum_{j \in \mathcal{U}_i}\mathbf{w}_{ij}^{H}\mathbf{w}_{ij}\leq P_{i}^{max},\ \forall i \in \mathcal{I}.
\end{eqnarray}
\end{subequations}
The optimization problem (\ref{EHM2}) is convex now, and can, therefore, be solved using CVX solver, e.g., Mosek \cite{aps2019mosek,andersen2000mosek}.
More specifically, the overall algorithm used to solve the original problem (\ref{EHM2}) first obtains the feasible solution by solving the minimum SINR maximization problem. The algorithm then determines the optimal solution of problem (\ref{EHM2}), and uses such a solution to convexify some of the non-convex constraints as highlighted above. The algorithm terminates when convergence is achieved, i.e., when the difference between two consecutive iterations falls below a predefined tolerance threshold~$\epsilon$.
The overall algorithm can then be summarized as follows:
\begin{algorithm}
 \caption{Proposed weighted sum rate maximization beamforming design algorithm }
 \label{G:SR}
 \begin{algorithmic}
  \State $\nu=0$, $\hat{\gamma}_{j}$ and $\mathbf{\hat{w}}_{ij}$ are feasible;
\While {$|\sum_{j=1\in \mathcal{U}}\hat{a}_j R_j^{v} -\sum_{j=1\in \mathcal{U}} \hat{a}_jR_j^{v-1}|> \epsilon$}
 \State $\bar{\gamma}_{j}, \mathbf{w}_{ij}=\arg \max \sum_{j=1\in \mathcal{U}}R_j(v)$;
\State $\{\hat{\gamma}_{j}^v, \mathbf{\hat{w}}_{ij}^{v}\}=\{\bar{\gamma}_{j}, \mathbf{w}_{ij}\}$;
\State $\nu=\nu+1$;
\EndWhile
\end{algorithmic}
\end{algorithm}
\par
\subsection{Computational Complexity}
The paper now analyzes the overall computational complexity of the proposed algorithms. The formulated GAP for the user association is solved using the BnB method, the complexity of which is approximately $O(\xi^n)$, where $1 < \xi < 2$, and $n$ represents the total count of variables $\alpha_{ij}$.
For the beamforming design, the complexity of each subproblem and the number of iterations in the loop determine the overall complexity. For the minimum SINR maximization problem (\ref{EHM_sinr}), we assume the number of loop iterations is $T_1$. Each subproblem (\ref{EHMsinBF1}) is formulated as a second-order cone program (SOCP) due to its quadratic convex objective and quadratic convex constraints,
which gives a computational complexity of $O(d_1^{3.5})$, where $d_1=2(N_U^{\text{HAPS}}N_A^{\text{HAPS}}+N_U^{\text{BS}}N_A^{\text{BS}})+1$ is the total number of variables, $N_U^{\text{HAPS}}$ is the number of users served by HAPS, and $N_U^{\text{HAPS}}$ is the number of users served by BSs. Therefore, the computational complexity of the Algorithm 1 is $O(\xi^n)+T_{1}O(d_1^{3.5})$. \par Similarly, for the weighted sum rate maximization problem (\ref{EHM}), we assume the number of iterations of loop is $T_2$. In addition, the computational complexity of each subproblem (\ref{EHM2}), which is solved by interior point methods is $O(n^{3.5}\log(1/\epsilon))$, where $\epsilon$ is the tolerance measurement. Therefore, the computational complexity of the overall algorithm is $O(\xi^n)+T_2 O(d_2^{3.5}\log(1/\epsilon))$, where $d_2=2(N_U^{\text{HAPS}}N_A^{\text{HAPS}}+N_U^{\text{BS}}N_A^{\text{BS}})+N_U$.

\subsection{Distributed Implementation}
This section outlines how we can implement Algorithms \ref{G:R_min} and \ref{G:SR} in a distributed manner across the multi-HAPS and ground BSs. Given the user association scheme, the set of users served by transmitter $i$ is $\mathcal{U}_{i}$. Define the local beamforming vector associated with transmitter $i$ as $\mathbf{w}_i=\operatorname{vec}\left(\left\{\mathbf{w}_{ij} \mid \forall j \in \mathcal{U}_i\right\}\right)$ and the corresponding introduced variables as $\bar{\boldsymbol{\gamma}}_{i}=\operatorname{vec}\left(\left\{\bar{\gamma}_{ij} \mid \forall j \in \mathcal{U}_i\right\}\right)$. Therefore, each transmitter $i$ only uses its own $\mathbf{w}_i$ with the interference terms caused by transmitter $b$ for the users served by transmitter $i$ ($\forall b\neq i$), which is denoted as $I_{ij}=\sum_{b\neq i }\sum_{u\in \mathcal{U}_{b}}|\mathbf{h}_{bj}^{H}\mathbf{w}_{bu}|^{2}, \forall j \in \mathcal{U}_i$.
\subsubsection{Maximization of Minimum $\mathrm{SINR}$}
Each transmitter $i$ aims to maximize the minimum $\mathrm{SINR}$ $\gamma_{\text{min}}^i$ among the served users.
Therefore, the function (\ref{Eq: g1}) can be written as:
\begin{equation}
\begin{aligned}
\hat{g}_1(\gamma_{\text{min}}^i, \hat{\gamma}_{\text{min}}^i, \hat{\mathbf{w}}_{i j}, \mathbf{w}_{i j})= & \sigma^2+ I_{ij} +\frac{\gamma_{\text{min}}^i}{(\hat{\gamma}_{\text{min}}^i)^2}|\mathbf{h}_{i j}^H \hat{\mathbf{w}}_{i j}|^2 \\
& -\frac{2}{\hat{\gamma}_{\text{min}}^i} \Re\left\{\hat{\mathbf{w}}_{i j}^H \mathbf{h}_{i j} \mathbf{h}_{i j}^H \mathbf{w}_{i j}\right\}.
\end{aligned}
\end{equation}
As a result, each transmitter can solve problem (\ref{EHMsinBF}) independently. Specifically, transmitter $i$
addresses the following problem locally:
\begin{subequations}
\label{EHM6}
\begin{eqnarray}
\label{EHM56}
&\displaystyle \max_{\mathbf{w}_{i}, \gamma_{\text{min}}^i}&  \gamma_{\text{min}}^i\\
\label{EHM6b}
&s.t.&(\ref{EHMsinrBFc})\\
\label{EHM6c}
 &&\hat{g}_1(\gamma_{\text{min}}^i, \hat{\gamma}_{\text{min}}^i, \hat{\mathbf{w}}_{i j}, \mathbf{w}_{i j}) \leq 0.
\end{eqnarray}
\end{subequations}
\subsubsection{Maximization of Weighted Sum Rate}
Observe that the  function $g_2$ can be written as:
\begin{eqnarray}
    \label{Fun:g2_hat}
    \hat{g}_2(\bar{\gamma}_{j}, \hat{\gamma}_{j},\mathbf{\hat{w}}_{ij}, \mathbf{w}_{ij} )&\!\!\!=\!\!\!&\sigma^{2} + I_{ij}
    + \frac{\bar{\gamma}_{j}}{(\hat{\gamma}_{j})^2} |\mathbf{h}_{ij}^{H}\mathbf{\hat{w}}_{ij}|^{2} \nonumber\\
    & &
    - \frac{2}{\hat{\gamma}_{j}}\Re\left\{\mathbf{\hat{w}}_{ij}^{H}\mathbf{h}_{ij} \mathbf{h}_{ij}^{H}\mathbf{w}_{ij} \right\}.
\end{eqnarray}
Hence each transmitter would be able to solve problems (\ref{EHM2}) locally. That is, each transmitter $i$ solves the following problem locally:
\begin{subequations}
\label{EHM5}
\begin{eqnarray}
\label{EHM5a}
&\displaystyle \max_{\mathbf{w}_{i}, \boldsymbol{\bar{\gamma}}_{j}}&
 \sum_{j\in \mathcal{U}_i} \hat{a}_jB\log_2(1+\bar{\gamma}_j),\\
\label{EHM5b}
&s.t.&(\ref{EHM2b}), (\ref{EHM2d}), \\
\label{EHM5c}
 &&\hat{g}_2(\bar{\gamma}_{j}, \hat{\gamma}_{j},\mathbf{\hat{w}}_{ij}, \mathbf{w}_{ij} ) \leq 0.
\end{eqnarray}
\end{subequations}
Compared to the centralized implementation of Algorithms \ref{G:R_min} and \ref{G:SR}, the above distributed algorithms implementation requires the following steps: (1) Each transmitter $i$ implements the algorithm independently and in parallel to solve the problems (\ref{EHM6}) and  (\ref{EHM5}), respectively. (2) Each transmitter checks for convergence individually. (3) After each iteration, the transmitters exchange the interference information $I_{ij}$. Consequently, each transmitter can update its beamforming vectors in parallel until convergence is achieved by all transmitters.
Such a highlight of our proposed multi-HAPS approach illustrates how Algorithms \ref{G:R_min} and \ref{G:SR} are amenable to distributed implementation, which underlines the numerical performance of the proposed algorithms within the scope of future multi-HAPS-ground networks.

\section{Simulation Results}
In this section, we present simulation results to assess the performance of the proposed algorithm across various network scenarios and parameter configurations. The below results, particularly, highlight the strong potential of multi-HAPS systems in promoting digital inclusion and supporting the democratization of future networks.
\subsection{Numerical Results}
\begin{table}[!t]
\centering
\caption{Simulation Parameters}
\label{tableI}
\begin{tabular}{|p{.34\textwidth} | p{.1\textwidth} |}
\hline
Parameter &  Value\\
\hline
  Bandwidth of BSs, $B$ & $10$ MHz  \\
  Rician factor, $\kappa_{HAPS}$ & $10dB$\  \\
  Ground BS antenna, $N_{A}^{BS}$& $3$ \\
  The number of HAPS antennas, $N_{A}^{\text{HAPS}}$& $50$\\
  The urban BS maximum power, $P_{BS,1}^{max}$& $1$w\  \\
  The suburban BS maximum power, $P_{BS,2}^{max}$& $2$w\  \\
  The rural BS maximum power, $P_{BS,3}^{max}$& $10$w\  \\
  The HAPS maximum power, $P_{HAPS}^{max}$& $100$w\\
  Standard deviation of ground-level shadowing $\sigma_{a}$& $6$dB\ \\
  \hline
\end{tabular}
\end{table}
Our simulations cover three distinct areas. Specifically, Area 1 consists of 30 BSs, located at coordinates \( x: (0 \, \text{km} \, \text{to} \, 5 \, \text{km}) \) and \( y: (0 \, \text{km} \, \text{to} \, 5 \, \text{km}) \), and serves 50\% of the overall user population. Area 2 includes 14 BSs, positioned at coordinates \( x: (25 \, \text{km} \, \text{to} \, 30 \, \text{km}) \) and \( y: (85 \, \text{km} \, \text{to} \, 90 \, \text{km}) \), and serves 30\% of the overall user population. The remaining area, Area 3, accounts for 20\% of of the overall user population.
Such a deployment is meant to infer that Area 1 is innately urban, Area 2 is suburban, and Area 3 is rural.
To enable comparisons, we consider four different scenarios. Scenario 1 deploys only the ground BSs as described above. Scenario 2 deploys a HAPS at the center of the rural area at coordinates [15, 45, 20] km.
Scenario 3 deploys two HAPSs, one in the center of the rural area and the other in the center of the suburban area at coordinates [27.5, 87.5, 20] km. Scenario 4 deploys three HAPSs, one in the rural center, one in the suburban center, and the third in the center of the urban area at coordinates [2.5, 2.5, 20] km.
For the sake of illustration, the data-availability variables $\beta_{ij}$ are assigned a value of 1 for the entire simulation section. Furthermore, we assume identical weights for all users by setting the associated weight \( a_{ij} = 1 \) for all \( i \in \mathcal{I}, j \in \mathcal{U} \). This assumption simplifies the analysis and enables us to evaluate the effects of resource allocation strategies without the influence of user-specific priority bias.
Consequently, the objective function reduces to maximizing the overall sum rate.
Table \ref{tableI} outlines the parameters employed in the simulation (unless mentioned otherwise). Other system level parameters can be found in \cite{liu2023joint}. Additionally, all simulations were performed in MATLAB using CVX, with Mosek 9.1.9 employed as the internal solver.
\par
\begin{figure}[!t]
\centering
\includegraphics[width=3in]{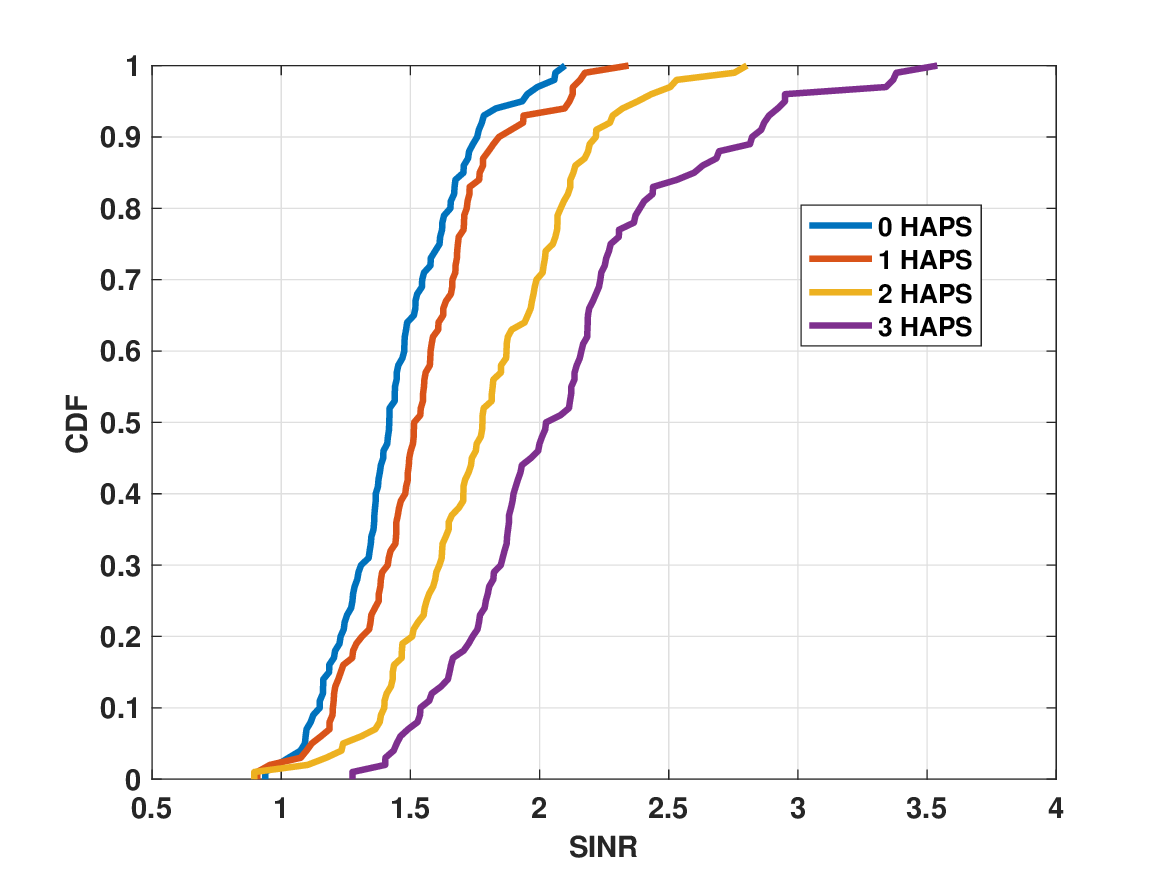}
\caption{CDF of minimum SINR.}
\label{Fig:sinr}
\end{figure}
We first plot the cumulative distribution function (CDF) of the SINR relying on maximizing the minimum SINR for different scenarios in Fig. \ref{Fig:sinr}, when the number of user is $60$. The results show that our proposed multi-HAPS NTN-ground integrated network significantly improves the SINR compared to standalone ground networks, particularly as the number of HAPSs increases. Such an enhancement stems from the ability of NTN to connect more users to the multiple HAPSs, which benefits from higher-order MIMO antennas and LoS links, resulting in enhanced SINR.
\par
\begin{figure}[!h]
\centering
\includegraphics[width=3in]{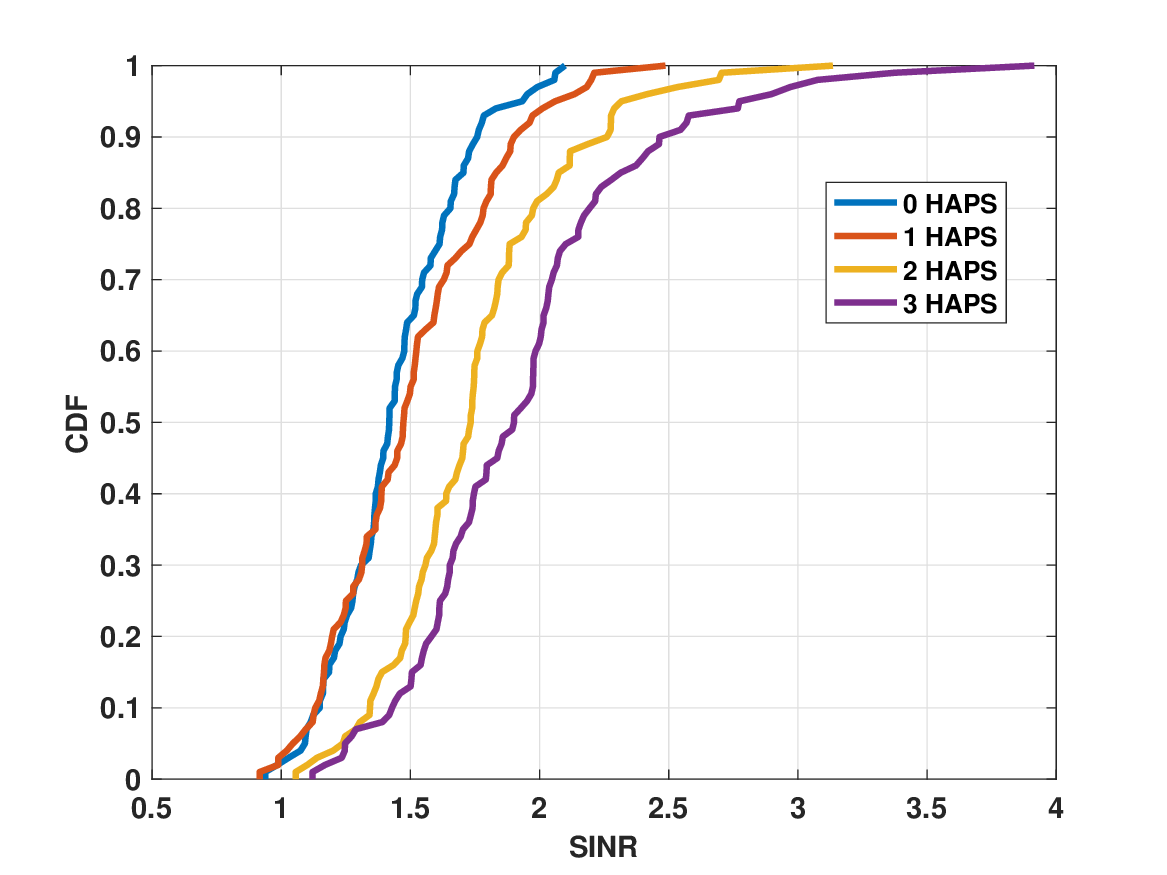}
\caption{CDF of minimum SINR with directional antenna}
\label{Fig:SINR_AN}
\end{figure}
The adoption of realistic antenna radiation patterns is essential for accurately characterizing system performance in terms of coverage, directional gain, and interference behavior. To reflect the impact of antenna directivity, we incorporate the elevation angle, defined as
$
\theta=\arctan(\frac{H}{\sqrt{(x-x_{HAPS})^2+(y-y_{HAPS})^2}}),
$
where $(x,y)$ and $(x_{HAPS},y_{HAPS})$ denote the azimuth coordinates of the user and the HAPS, 
respectively, and $H$ represents the altitude of the HAPS.
In our model, we impose directional gain constraints based on the elevation angle to account for the limited beamwidth of HAPS antennas. Specifically, for rural deployments, if the elevation angle exceeds 
$60^{\circ}$, the antenna gain is set to $-30$dB, reflecting significant attenuation outside the main lobe. For suburban and urban environments, a more stringent constraint is applied, where an elevation angle exceeding $10^{\circ}$, results in the same penalty of $30$dB. 
This approach approximates the practical beam coverage limitations of high-gain, narrow-beam HAPS antennas.
 Figure~\ref{Fig:SINR_AN} presents the cumulative distribution function (CDF) of the SINR obtained by maximizing the minimum SINR across different scenarios incorporating directional gain constraints. It can be observed that the design utilizing directional antenna gains generally achieves superior SINR performance compared to the scenario with omnidirectional antennas shown in Figure~\ref{Fig:sinr}. This improvement arises from the narrower beamwidth of directional antennas, which effectively reduces interference by focusing energy towards the intended users. 

\par
\begin{figure}[!t]
\centering
\includegraphics[width=3in]{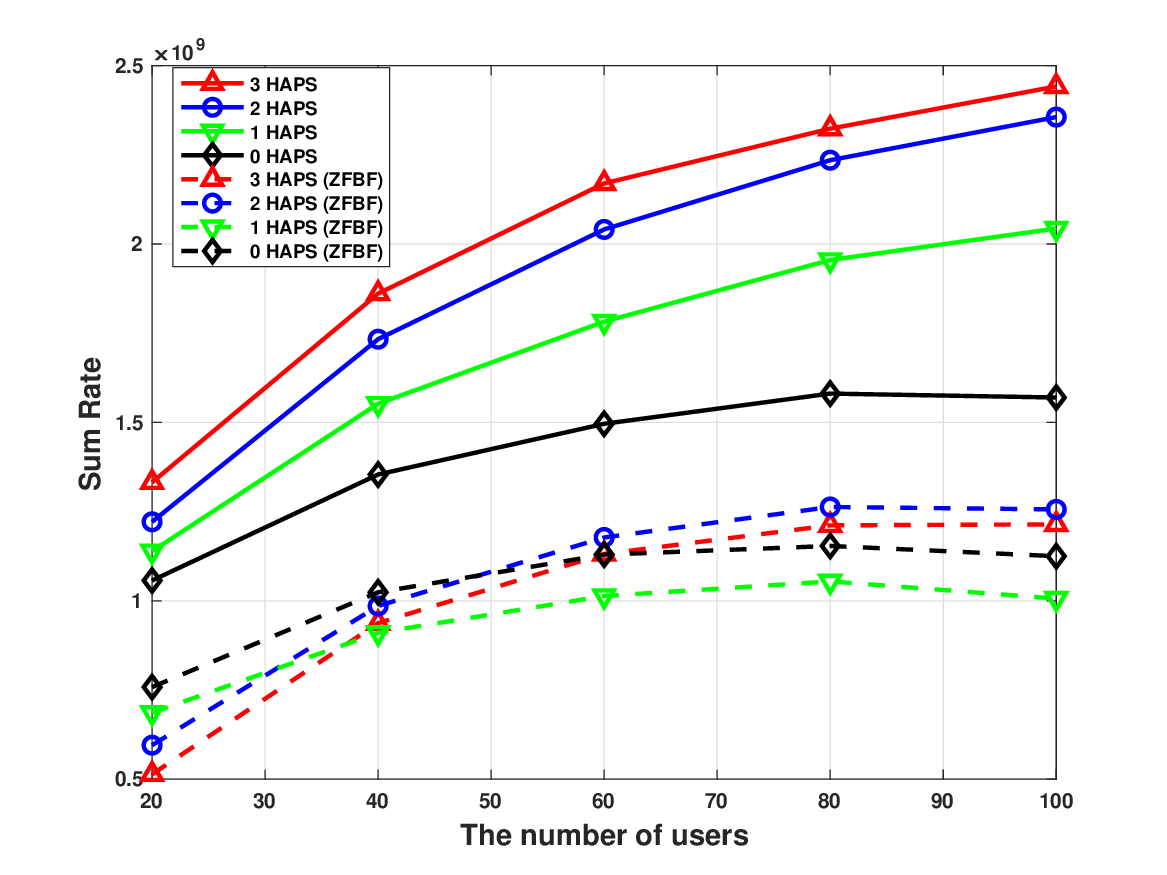}
\caption{Sum rate versus the total number of users.}
\label{Fig:user_sum_rate}
\end{figure}
The effect of the total number of users on the sum rate is subsequently examined, as depicted in Fig. \ref{Fig:user_sum_rate}, where the minimum SINR for each user $\gamma_{j,\text{min}}$ is set to $0.5$, i.e., part of the feasibility guarantee step. Specifically, as illustrated in Fig. \ref{Fig:user_sum_rate}, the overall network sum rate exhibits an increasing trend with the rise in the number of users. Furthermore, the figure particularly highlights how deploying additional HAPSs enhances the system throughput across different numbers of users. This improvement is attributed to the proposed algorithm capabilities as a powerful interference management technique, which enables HAPS to complement the capabilities of ground-level BSs by mitigating interference, thereby serving a larger number of users more effectively. 
To further highlight the performance of our proposed algorithms, we now add the zero-forcing beamforming (ZFBF) as an additional beamforming design baseline. In this context, ZFBF is implemented on a per-transmitter basis, i.e., it helps eliminating the intra-mode interference \cite{yoo2006optimality}, namely, intra-HAPS interference, and intra-BS interference in the current paper setup. Fig.~\ref{Fig:user_sum_rate}  illustrates the sum rate performance versus the total number of users. As the number of users increases, the sum rate under ZFBF initially improves due to its ability to spatially separate users within the same transmission. However, ZFBF does not address inter-mode interference (i.e., interference between different transmitters such as HAPS and BSs), which becomes more significant in dense user scenarios. Consequently, the performance of ZFBF saturates and eventually declines as interference intensifies.
In contrast, our proposed method consistently achieves superior performance across all user densities. This improvement stems from our proposed algorithm reliance on joint interference management strategy, which effectively mitigates both intra-mode and inter-mode interference.
\par
\begin{figure}[!t]
\centering
\includegraphics[width=3in]{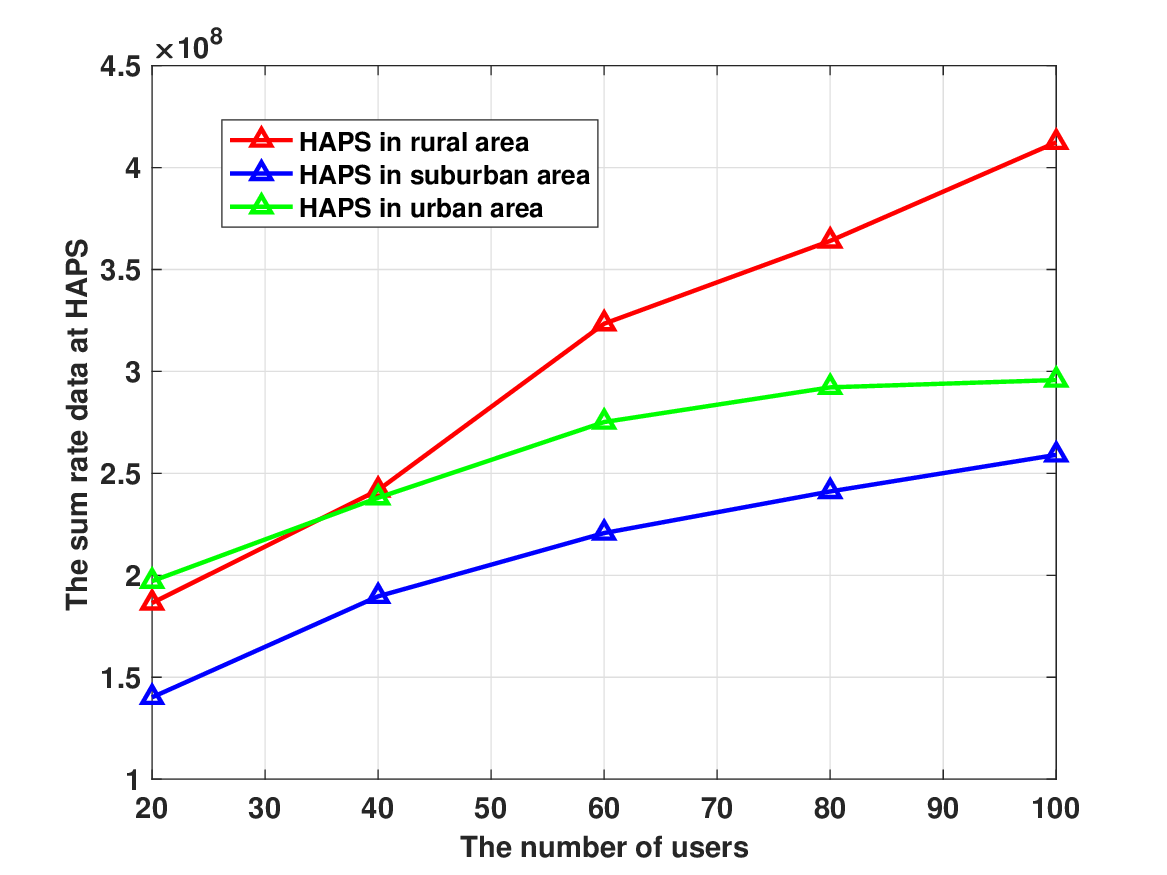}
\caption{Sum rate at HAPS in differnt area versus the total number of users.}
\label{Fig:user_sum_rate_HAPS}
\end{figure}
The throughput capability of a HAPS is constrained by factors such as the maximum available bandwidth, limited processing resources, and computational capacity. To this end, Fig.~\ref{Fig:user_sum_rate_HAPS} illustrates the sum rate at the HAPS versus the number of users under three different scenarios (i.e., rural, suburban, urban) in a three-HAPS setup. Fig.~\ref{Fig:user_sum_rate_HAPS} particularly shows that the sum rate at the HAPS increases with the number of users. Moreover, the throughput of the HAPS in rural areas is the highest. This is because rural regions have fewer ground base stations (GBSs) and thus rely more heavily on HAPS as user density grows. In contrast, the throughput growth in urban and suburban areas is slower due to the dense and well-established deployment of GBSs. It is noteworthy that when the HAPS throughput capacity exceeds approximately 450 Mbps, the throughput limitation becomes inactive. This assumption is reasonable, as such a capacity level aligns with the practical performance range of current HAPS systems.
\par
\begin{figure}[!h]
\centering
\includegraphics[width=3in]{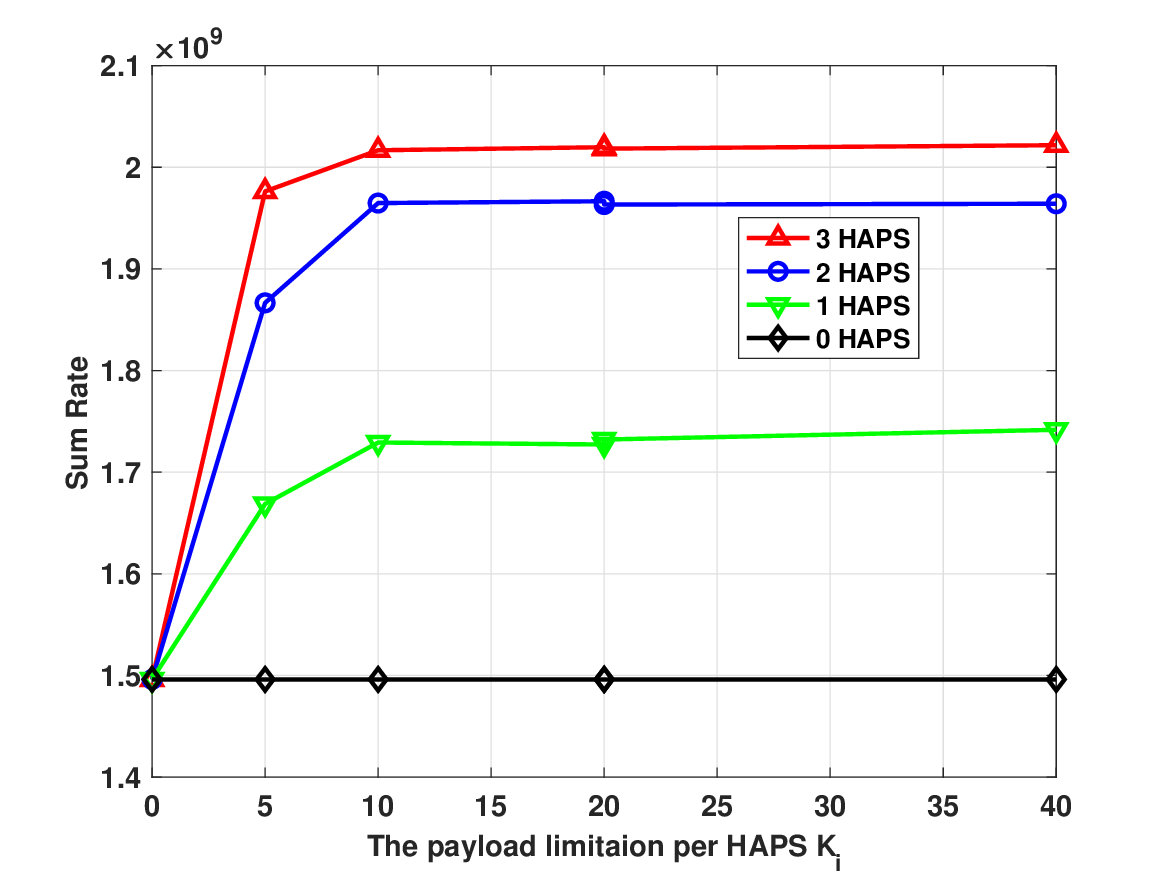}
\caption{Sum rate versus payload limitation per HAPS $K_i$.}
\label{Fig:K_sum_rate}
\end{figure}
To better reflect on the general impact of the considered payload constraint, we assume that each HAPS $i$ is capable of serving at most 
$K_i$ users, where $K_i$ represents the payload limitation of HAPS $i$. More specifically, we now add Fig.~\ref{Fig:K_sum_rate} which presents the  sum rate performance as a function of $K_i$, where the total number of users is set to $60$ and the number of antennas per HAPS is set to $40$. It can be observed that the sum rate increases with $K_i$ and eventually saturates, indicating that the system reaches its maximum serving capacity. Furthermore, deploying additional HAPSs leads to a higher sum rate, demonstrating that HAPSs can effectively enhance the capacity of the ground network. 
\par
\begin{figure}[!t]
\centering
\includegraphics[width=3in]{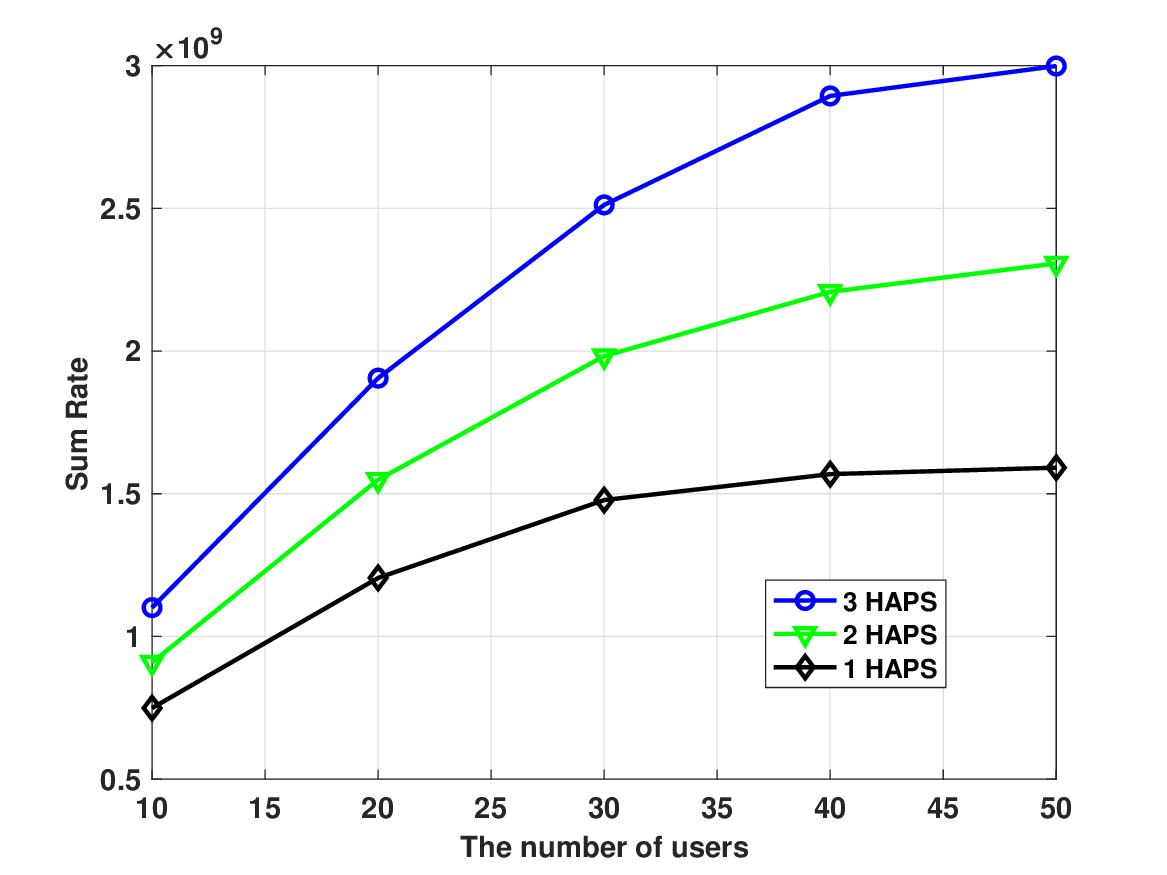}
\caption{GBSs absence scenario: Sum rate versus the total number of users}
\label{Fig:user_sum_rate_BS0}
\end{figure}
To further evaluate the performance of HAPSs and better demonstrate their capability to support users in areas where terrestrial infrastructure is unavailable. 
Specifically, Fig.~\ref{Fig:user_sum_rate_BS0} illustrates how the sum rate varies with the total number of users under different numbers of HAPS in the absence of GBSs. Due to the capacity limitations of HAPS, the maximum number of users in this scenario is set to 50.
As shown in the figure, the sum rate increases with the number of users, indicating that the system can support a growing users even in the absence of GBSs. Moreover, it can be observed that the performance gap between different numbers of HAPS is significantly larger in the GBS-absent scenario compared to the case where GBSs are active. This indicates that, in the absence of GBSs, deploying additional HAPS becomes even more critical to maintain acceptable system performance.

\par
\begin{figure}[!t]
\centering
\includegraphics[width=3in]{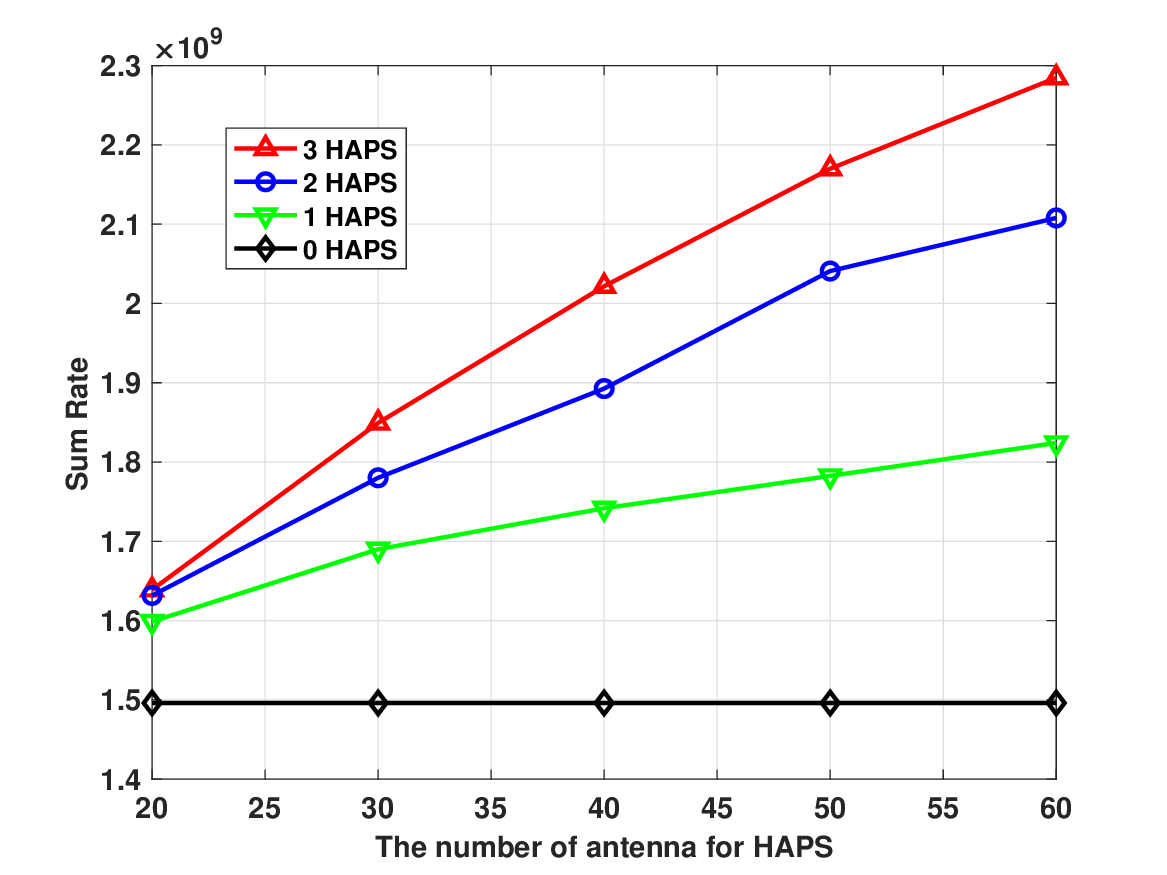}
\caption{Sum rate versus the number of HAPS antennas.}
\label{Fig:AN}
\end{figure}
Fig. \ref{Fig:AN} shows the impact of the number of HAPS antennas on the network performance when the number of users is 60 users. The figure highlights the performance gains achieved by integrating HAPS capabilities into the ground network, shown by the numerical benefit provided by the incremental addition of HAPSs. Moreover, with the increase in the number of HAPS antennas, the network's sum rate improves. It is also worth noting how Fig.~\ref{Fig:AN} illustrates that the sum rate for the BS-only scenario remains constant, as it is unaffected by the number of HAPS antennas.
\par
\begin{figure}[!t]
\centering
\includegraphics[width=3in]{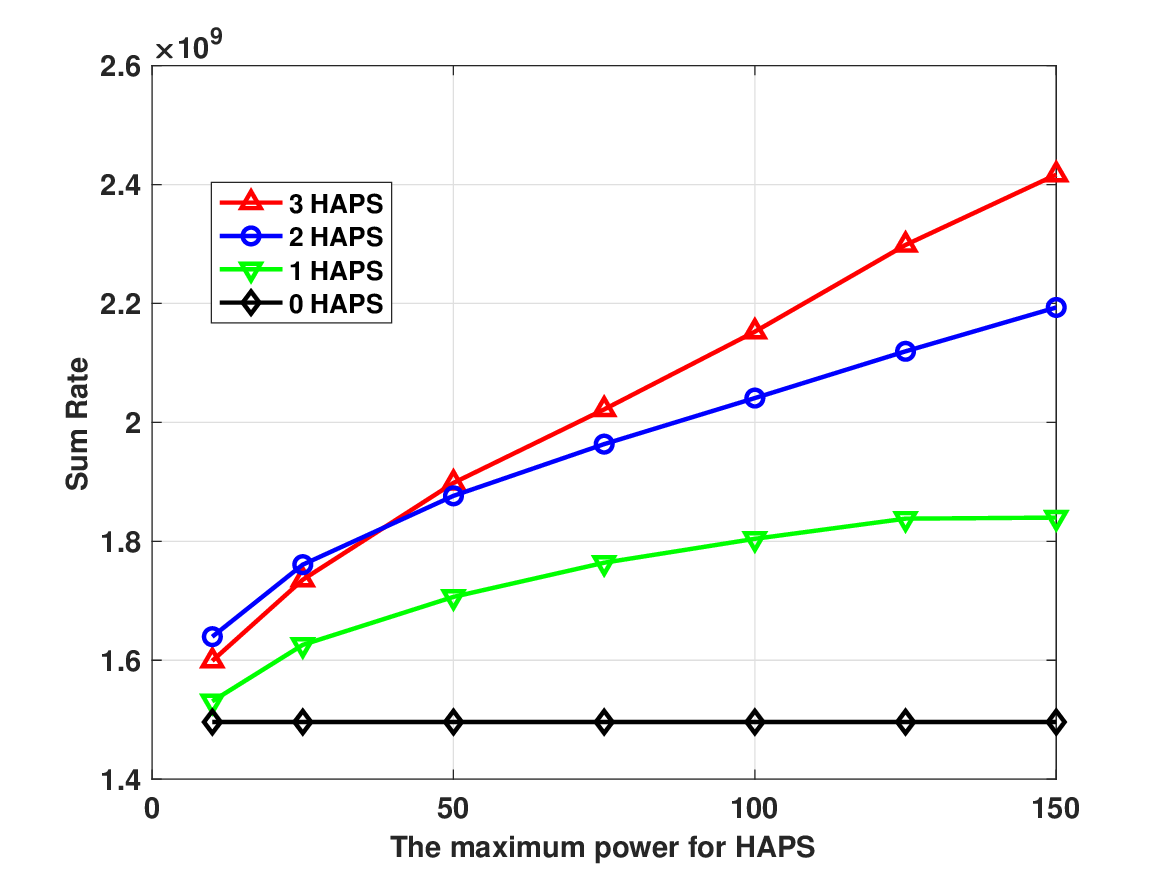}
\caption{Sum rate versus the maximum power of HAPS.}
\label{Fig:power}
\end{figure}

\begin{figure}[!t]
\centering
\includegraphics[width=3in]{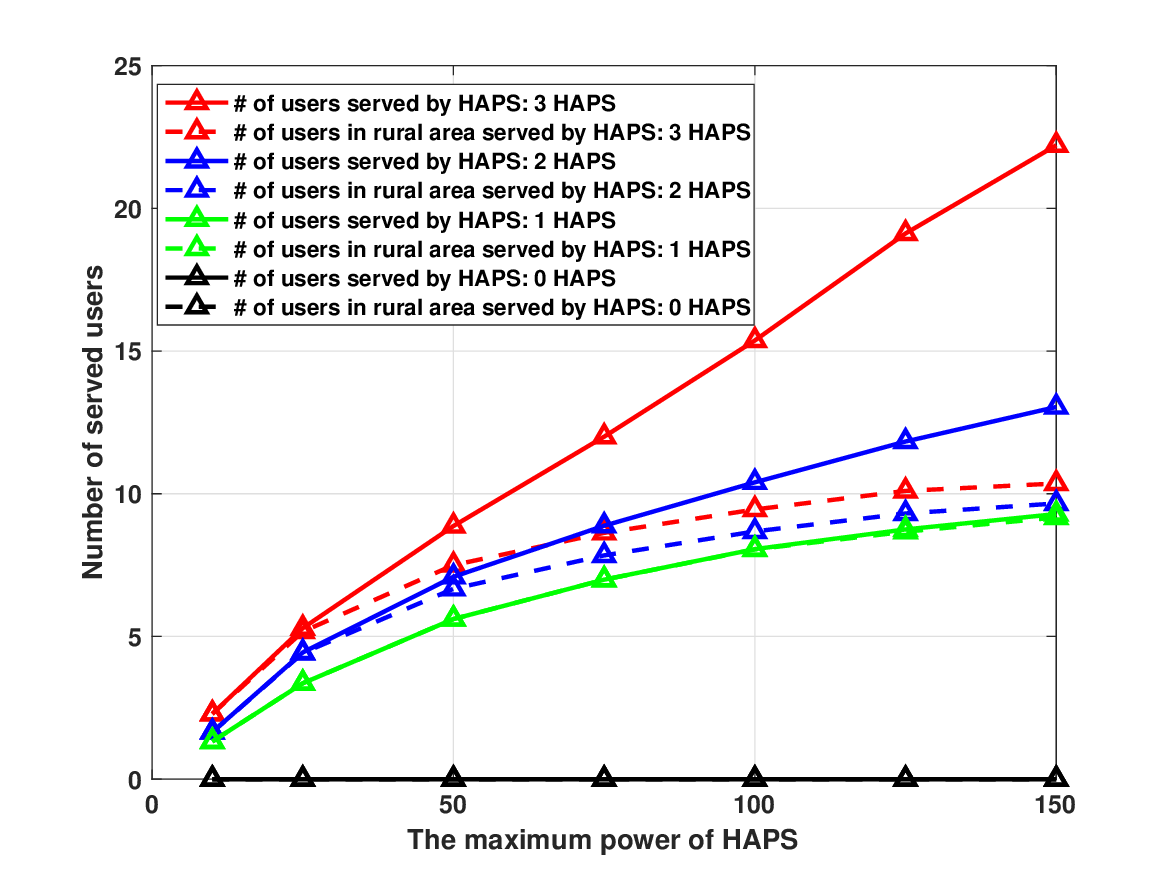}
\caption{The number of users served by HAPS with different HAPS power level.}
\label{Fig:user connection for different power level}
\end{figure}
Fig. \ref{Fig:power} illustrates the sum rate as a function of the maximum HAPS power, with the number of users fixed at 60.
The figure shows how the sum rate increases as the maximum power of HAPS grows. Notably, the gap between the 3-HAPS scenario and other scenarios becomes more pronounced at higher power levels, due to the increased number of connected users, as further verified through Fig. \ref{Fig:user connection for different power level}.
However, when the HAPS operates at a low maximum power level (e.g., 10W and 25W), the performance of the 2-HAPS scenario surpasses that of the 3-HAPS scenario. In fact, as illustrated in Fig. \ref{Fig:user connection for different power level}, at low power levels, all HAPSs primarily serve users in rural areas, leading to interference for users in suburban and urban areas. This is further due to the fact that, at such a low HAPS power level, the higher density of urban areas causes the 3-HAPS scenario to experience greater interference compared to the 2-HAPS scenario. An interesting takeaway message of Fig. \ref{Fig:user connection for different power level} is that with an increase in the maximum power of HAPS, there is a corresponding rise in the number of users that can be served by the HAPS, which illustrates the important role of HAPS in improving the overall digital inclusion.
\par
\begin{figure}[!t]
\centering
\includegraphics[width=3in]{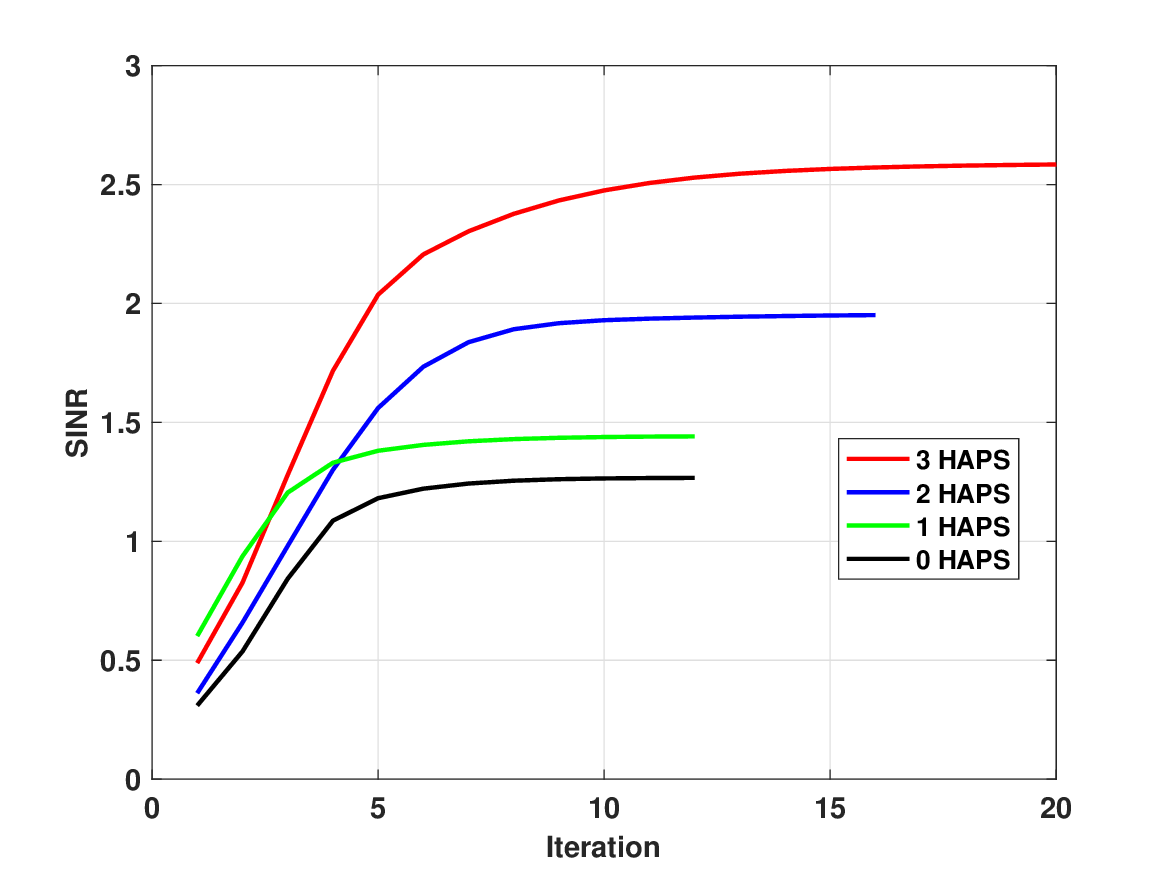}
\caption{Convergence behaviour of the minimum SINR maximization.}
\label{Fig:SINR_converge}
\end{figure}
Fig. \ref{Fig:SINR_converge} depicts the numerical convergence of Algorithm 1 by illustrating the progression of the minimum SINR values over successive SCA iterations for four distinct network configurations. The results show that the objective function value consistently increases with each iteration, stabilizing after five or six iterations in the 0 HAPS and 1 HAPS scenarios, and after eight or nine iterations in the 2 HAPS and 3 HAPS cases.
\par
\begin{figure}[!t]
\centering
\includegraphics[width=3in]{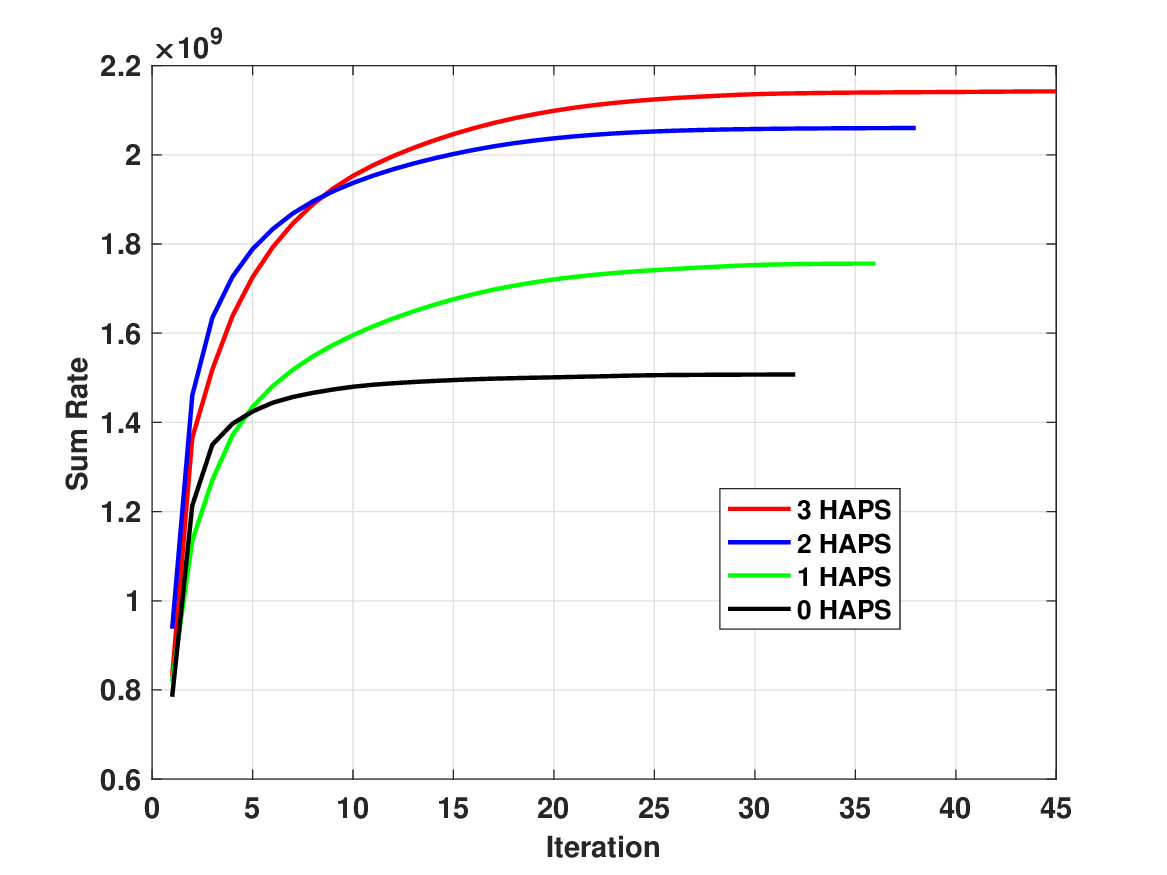}
\caption{Convergence behavior of the sum rate maximzation.}
\label{Fig:converge}
\end{figure}
Similarly, the convergence behavior of the proposed algorithm \ref{G:SR} is illustrated in Fig. \ref{Fig:converge} for various numbers of HAPSs serving $60$ users in total. Fig.~\ref{Fig:converge}, particularly, demonstrates that the overall algorithm converges at a reasonably fast rate for different scenarios, further highlighting the numerical effectiveness of the proposed approach.
\par
\begin{table*}[!h]
\centering
\caption{Computational time performance}
\label{table-time}
\begin{tabular}{|p{.32\textwidth} | p{.2\textwidth} | p{.2\textwidth} | p{.2\textwidth} | }
\hline
  \textbf{Schemes (max-min SINR) } &  \textbf{$N_U=10$}& \textbf{$N_U=20$}\\
 \hline
0 HAPS &16.9673s& 49.5465s \\
  \hline
1 HAPS& 21.2802s & 47.8494s \\
  \hline
2 HAPS & 21.4747s& 56.2421s\\
  \hline
3 HAPS& 31.1556s& 80.1044s\\
  \hline
\textbf{Schemes (max sum rate) } &  \textbf{$N_U=10$}& \textbf{$N_U=20$}\\
  \hline
0 HAPS &101.7993s& 399.6130s \\
  \hline
1 HAPS& 97.8046s& 454.2410s\\
  \hline
2 HAPS & 94.9995s &461.4635s\\
  \hline
3 HAPS& 108.6165s &486.1502s\\
\hline
\end{tabular}
\end{table*}
To address the computational complexity more concretely, Table \ref{table-time} presents the running times for solving both the minimum SINR maximization problem and the sum rate maximization problem under varying numbers of HAPSs and users.
The results are obtained
using MATLAB R2022b based on Intel(R) Xeon(R) W-2145 CPU @ 3.70GHz processor.
As observed in the table, the runtime of the algorithms increases with the number of users and HAPSs, which is consistent with the expected growth in computational complexity. Moreover, the results indicate that solving the sum rate maximization problem generally requires more computational time compared to the minimum SINR maximization problem. This is primarily because the latter serves as a preprocessing step for the former, and its solution is used to initialize or guide the optimization of the sum rate objective.
\par
\begin{figure}[!t]
\centering
\includegraphics[width=3in]{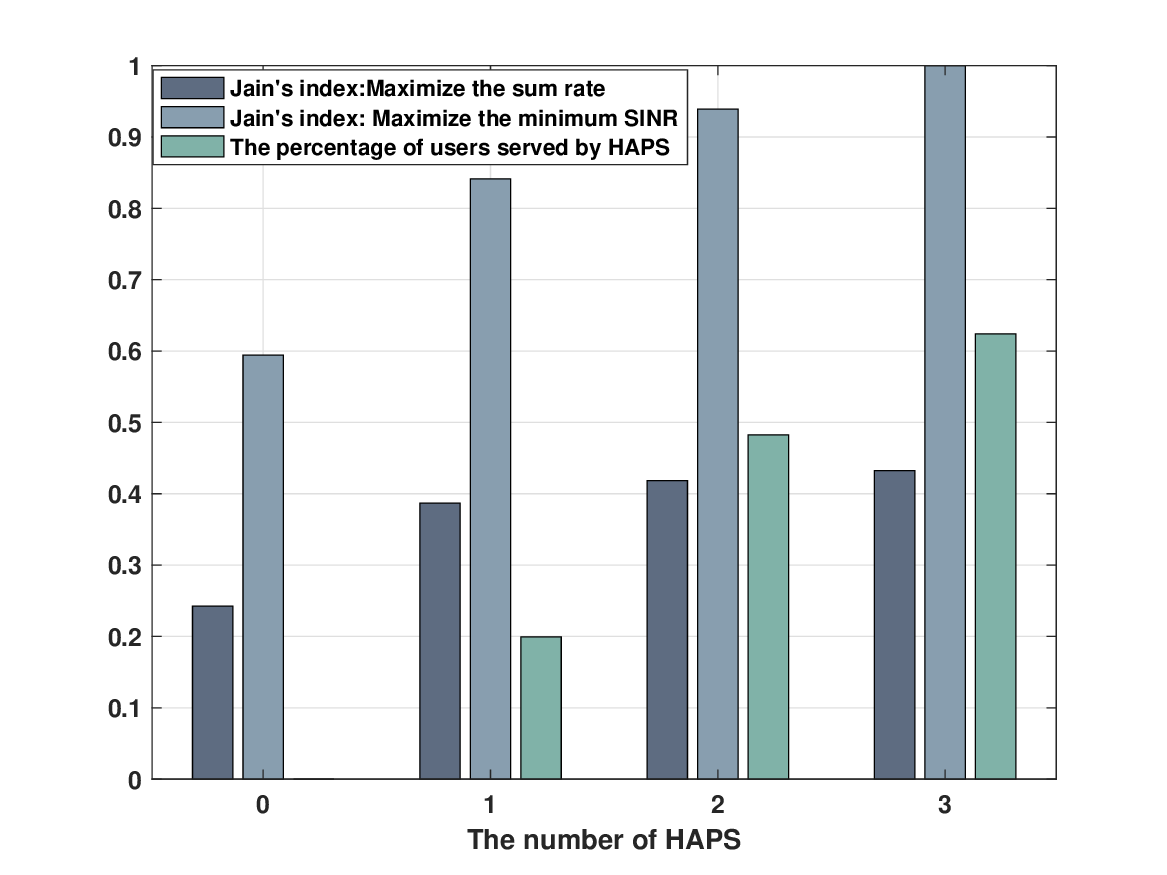}
\caption{Performance within different number of HAPS.}
\label{Fig: J_index}
\end{figure}
\par
\begin{figure}[!t]
\centering
\includegraphics[width=3in]{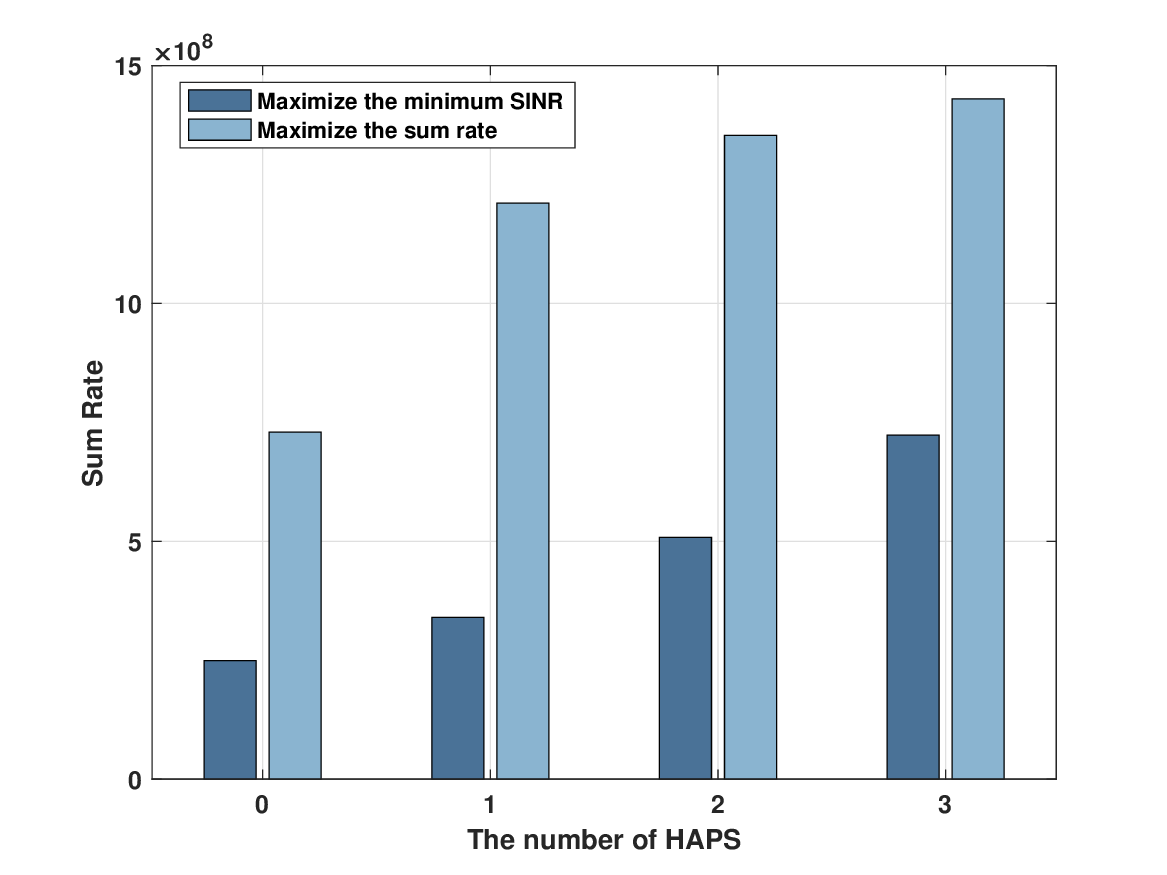}
\caption{Sum rate within different number of HAPSs.}
\label{Fig: sum_rate_SINR}
\end{figure}

\par
To further illustrate the digital inclusion prospects of the proposed framework, the fairness of users' rates is then assessed using Jain’s fairness index $J \in\left[\frac{1}{U}, 1\right]$, defined by the equation $J=\frac{\left(\sum_{N_U} R_i\right)^2}{N_U \sum_{N_U} R_i^2}$, where $U$ is the number of users and $R_i$ the rate for user $i$. In addition, the number of BS antennas is set to $1$, and the number of HAPS antennas is set to $20$. In BS only scenario, we assume that rural area users are unconnected. Fig.~\ref{Fig: J_index} demonstrates a notable improvement in Jain's fairness index when the number of HAPS increases.
This trend highlights the capability of HAPS to augment ground base stations.
The figure additionally illustrates that as the number of HAPSs increases, a larger proportion of users are allocated to HAPSs. This observation underscores the effectiveness of the proposed multi-HAPS approach in providing a democratized, digitally sustainable NTN connectivity platform. Furthermore, the results indicate that maximizing the minimum SINR achieves better fairness (higher Jain's index) compared to maximizing the sum rate. However, as shown in Fig.~\ref{Fig: sum_rate_SINR}, the sum rate is higher when prioritizing sum rate maximization over minimum SINR maximization.
\par
\begin{figure*}[!h]
\centering
\subfigure[BS only scenario]{
    \includegraphics[width=2.5in]{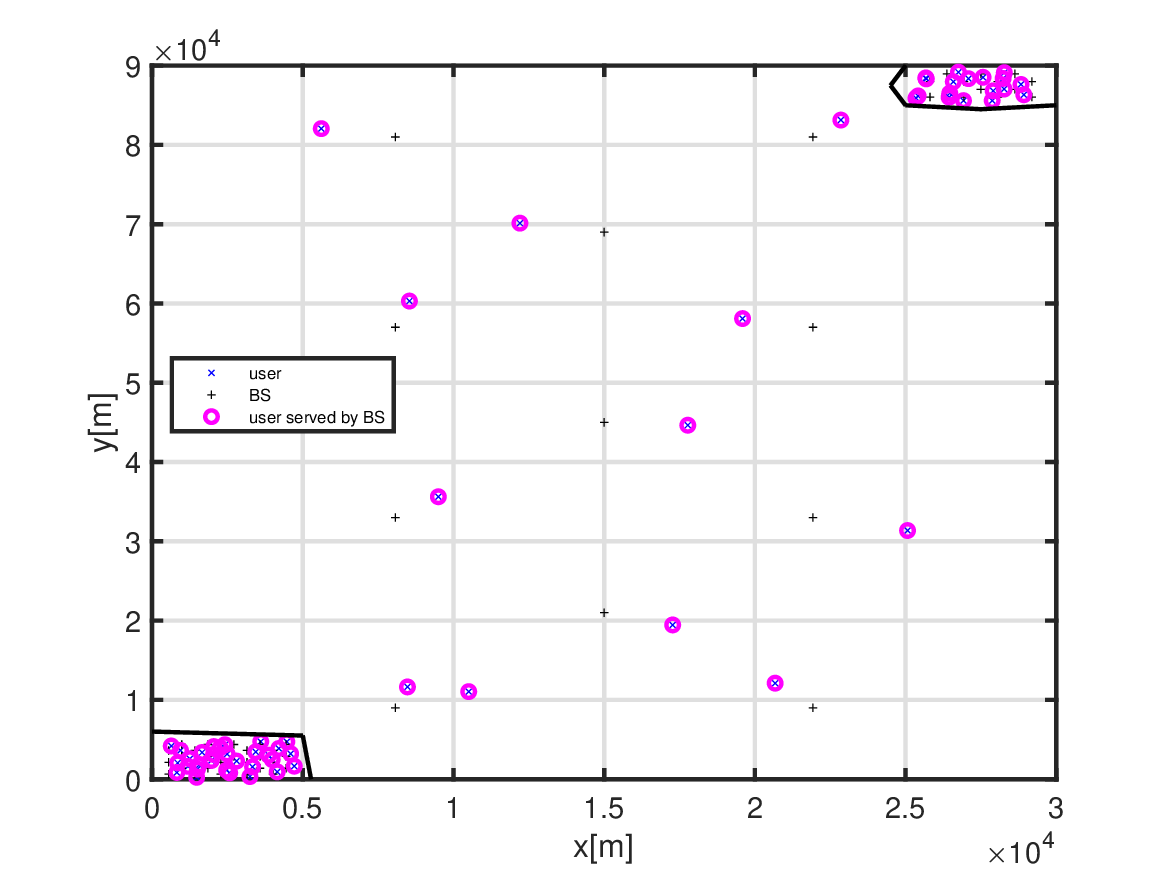}
    \label{Fig: BS}
}
\subfigure[1-HAPS scenario]{
    \includegraphics[width=2.5in]{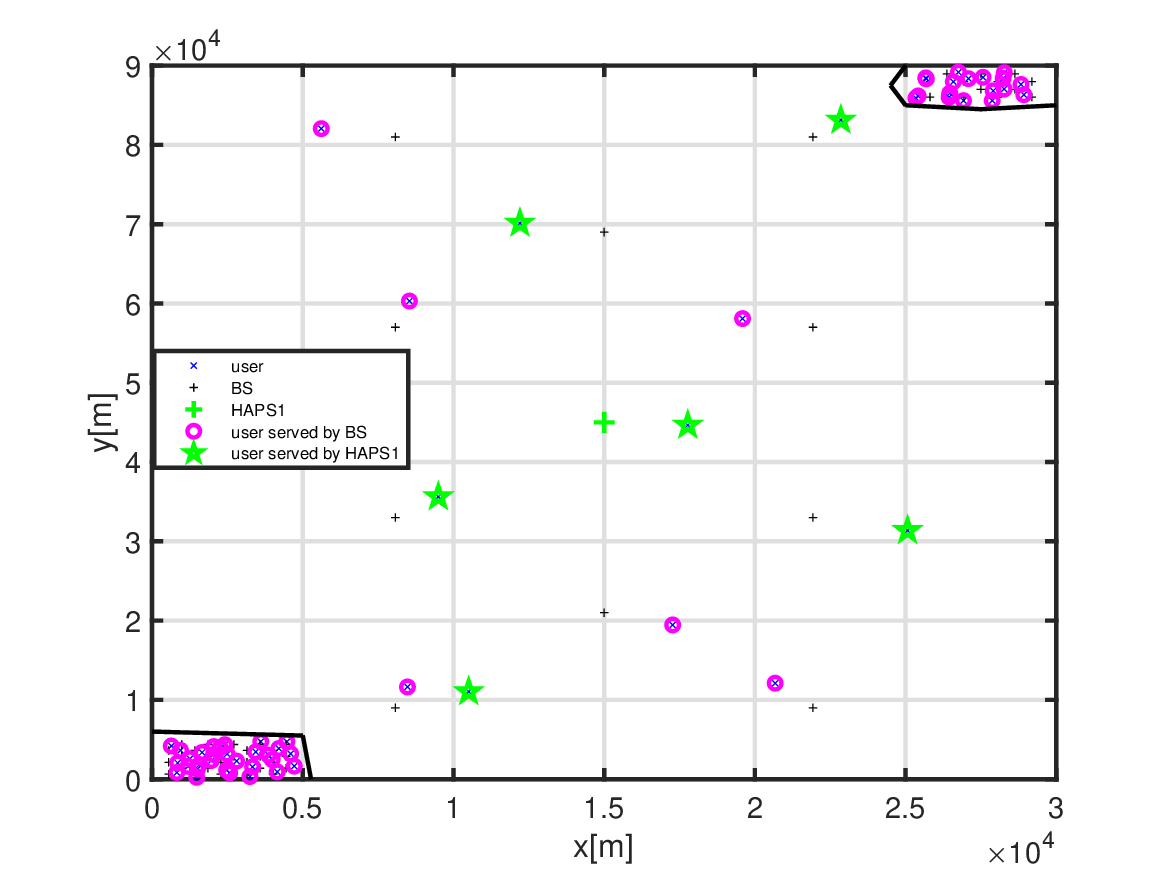}
    \label{Fig: 1HAPS}
}
\subfigure[2-HAPS scenario]{
    \includegraphics[width=2.5in]{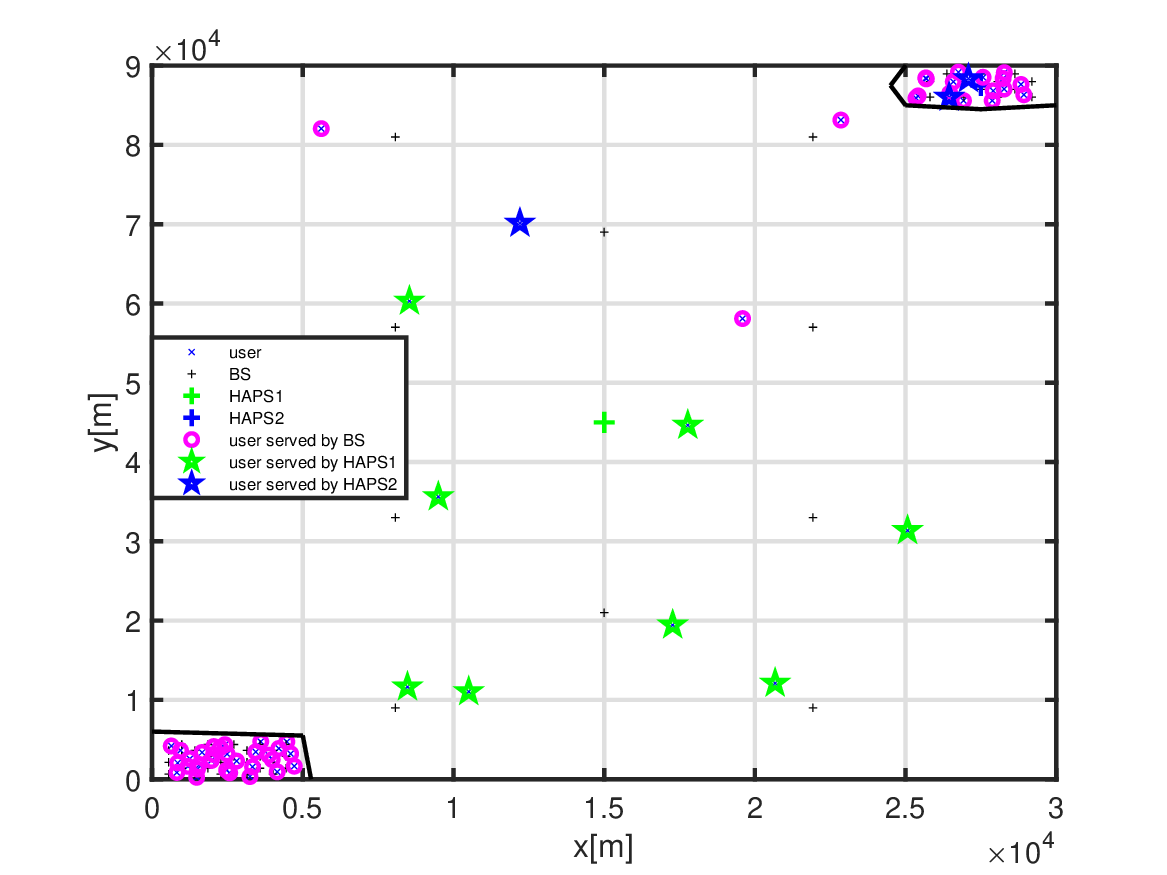}
    \label{Fig: 2HAPS}
}
\subfigure[3-HAPS scenario]{
    \includegraphics[width=2.5in]{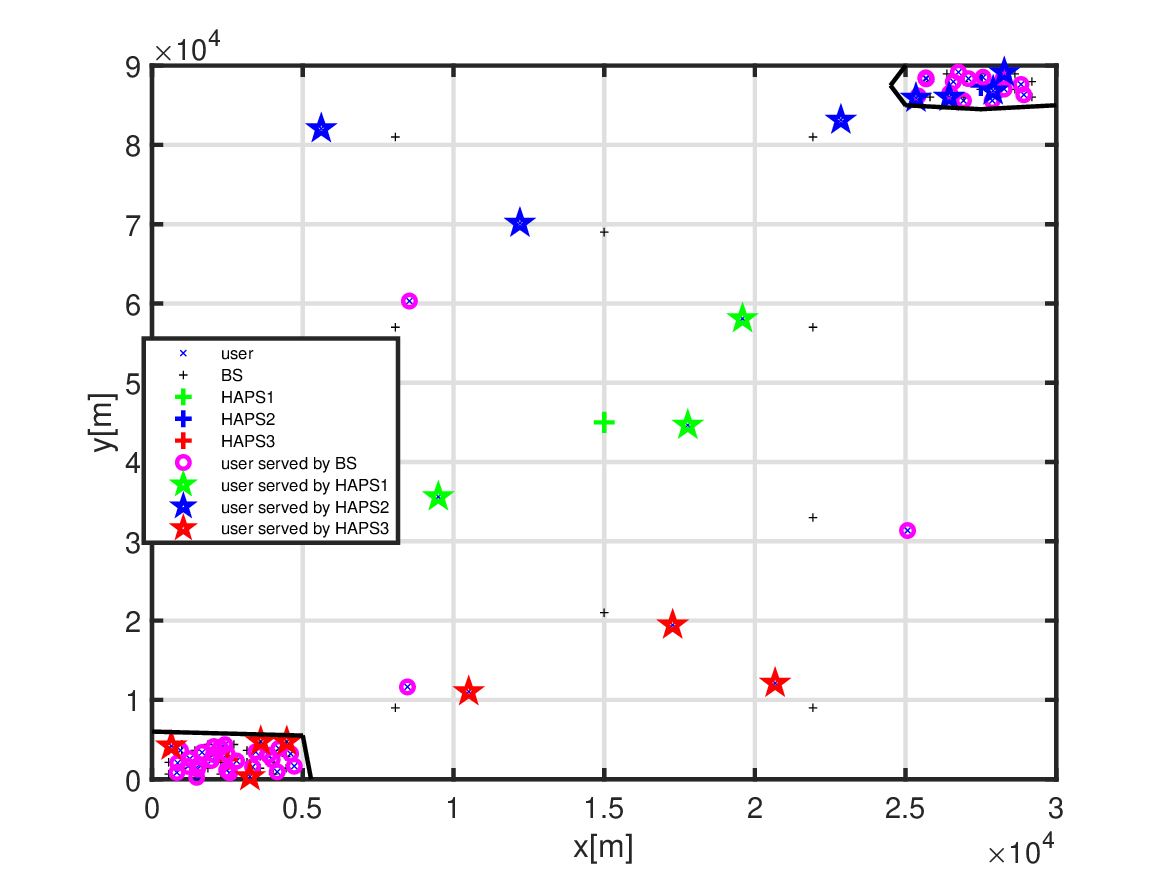}
    \label{Fig: 3HAPS}
}
\caption{User-to-HAPS and user-to-ground BS association for different numbers of HAPSs.}
\label{Fig: how_users_are_served}
\end{figure*}
Lastly, Fig.~\ref{Fig: how_users_are_served} offers a graphical depiction of the evolution of ground user associations as the number of HAPSs increases.
In the BS-only scenario, all users are served exclusively by ground BSs. In the 1-HAPS scenario, the HAPS begins assisting users in rural areas, demonstrating its capability to connect the unconnected. For the 2-HAPS scenario, the HAPS extends its coverage to serve users in both rural and suburban areas. Finally, in the 3-HAPS scenario, the HAPSs support users across the entire network, including urban areas.
Such results pertaining to the 2-HAPS and 3-HAPS scenarios particularly emphasize the role of the multi-HAPS augmented network in superconnecting the connected, thereby showcasing their ability to enhance future networks performance and coverage.
\subsection{Discussion and Recommendation}
The results presented in this paper demonstrate that, when compared to the traditional BSs-only scenario, the proposed multi-HAPS-assisted ground network offers significant improvements to both the minimum SINR among users and the overall sum rate performance of the system. This is especially the case when the HAPSs capabilities are enhanced (e.g., an increased number of antennas and elevated transmit power), as HAPS can support both unconnected and connected users, thus enhancing network fairness while contributing to overall digital sustainability and inclusion. One of this paper focuses is the per-user rate requirement, which ensures that all users remain connected and their QoS is guaranteed. This requirement, combined with advancements in HAPS mega-constellation design, makes the results particularly relevant in high-demand communication environments, such as large-scale events (e.g., concerts and sports events), where network congestion and performance challenges are critical. Furthermore, the results of this paper provide valuable insights into the potential for HAPS to play a pivotal role in advancing future 6G systems, particularly by boosting and complementing the performance of ground-level communication.

\section{Conclusion}
The agenda for 6G and beyond networks prioritizes fostering digital inclusion through the achievement of higher data rates and the assurance of QoS. Given the limitation of ground networks, future VHetNets promise to boost-up the digital inclusion of future wireless networks.
To this end, our paper proposes a multi-HAPS-ground integrated network, especially introduced to achieve the highly sought-after digital equity. Specifically, this paper focuses on optimizing user scheduling and designing the corresponding beamforming vectors to maximize the minimum SINR among all users, and to maximize the system weighted sum rate, constrained by HAPS and BS transmit power constraints, HAPS payload constraints, and minimum rate requirements. This paper addresses the inherent numerical complexity of both problems using GAP-type solution for the discrete association parts, and employing SCA to handle the beamforming design parts. The simulation results presented in the paper underscore the potential of the proposed multi-HAPS system in enhancing both network fairness and throughput, which promises to spearhead future research avenues pertaining to digitally sustainable large-scale terrestrial and non-terrestrial networks.  
In future work, we aim to integrate more realistic antenna models that account for 3D beam patterns, side-lobe characteristics, and adaptive beam steering mechanisms, thereby enabling more accurate system performance evaluation and optimization.
\par
Finally, future research directions of our work include incorporating the HAPS throughput limitations resulting from the maximum available bandwidth and limited processing power into our optimization framework, devising an optimized mechanism to determine the payload constraints, investigating energy-efficiency optimization to ensure system sustainability, considering imperfect channel information at both the HAPS and ground base stations, and integrating data-driven techniques to enhance the overall problem-solving framework.


\bibliography{my_bibliography}
\bibliographystyle{IEEEtran}

\end{document}